\documentstyle[12pt,psfig]{article}

\begin{document}

\vskip 1truecm
\rightline{Preprint PUP-TH-1544 (1995) and LANCS-TH/9517 (1995)}
\rightline{ e-Print Archive: hep-ph/9506475}
\bigskip
\centerline{\Large  How fast can the wall move?}
\medskip
\centerline{\it A study of the electroweak phase transition dynamics}
\bigskip
\centerline{\Large Guy D. Moore\footnote{e-mail:
guymoore@puhep1.princeton.edu}
}
\medskip

\centerline{\it Princeton University}
\centerline{\it Joseph Henry Laboratories, PO Box 708}
\centerline{\it Princeton, NJ 08544, USA}

\medskip

\centerline{ and }

\medskip

\centerline{\Large Tomislav Prokopec\footnote{e-mail:
T.Prokopec@lancaster.ac.uk,  address from Sep 1 '95: Cornell University}
}
\medskip

\centerline{\it Lancaster  University}
\centerline{\it School of Physics and Chemistry}
\centerline{\it Lancaster LA1 4YB, UK}

\medskip

\centerline{\bf Abstract}

\smallskip
We consider the dynamics of bubble growth in the Minimal Standard Model
at the electroweak phase  transition and determine the shape and the velocity
of the phase boundary, or bubble wall. We show that in the semi-classical
approximation the friction on the wall arises from the deviation of massive
particle populations from thermal equilibrium. We treat these  with
Boltzmann equations in a fluid approximation. This approximation is
reasonable for the top quarks and the light species while it underestimates the
friction from the infrared $W$ bosons and Higgs particles.  We use the
two-loop finite temperature effective potential and find a subsonic bubble wall
for the whole range of Higgs masses $0<m_H<90$GeV. The result is weakly
dependent on $m_H$: the wall velocity $v_w$ falls in the range $0.36<v_w<0.44$,
while the  wall thickness is in the range
$29> L T > 23 $. The wall is thicker than
the phase equilibrium value because out of equilibrium particles exert
more friction on the back than on the base
of a moving wall.  We also consider the effect of an infrared gauge condensate
which may exist in the symmetric phase; modelling it simplemindedly, we
find that the wall may become supersonic, but not ultrarelativistic.

\section{Introduction}

It has gradually become clear that the baryon asymmetry of the universe
may have been created at a first order electroweak phase
transition.  The most important ingredient, baryon number violation, was
demonstrated in the Standard Model by t'Hooft \cite{tHooft} and was later
shown to proceed rapidly at high temperatures \cite{Rubakov,ArnoldMac}.
The necessary departure from thermal equilibrium is  supplied by the
first order electroweak phase transition \cite{Krizhnits},
 which has recently been the subject of intense investigation
\cite{DineLinde,Arnold,Fodor,Shapnonpert,Kajantie,Wetterich,Fodor2}. There have
also been promising developments in understanding the mechanisms by which
baryogenesis may proceed.

The general picture is this: at high temperatures, such as those prevalent
in the early universe, thermal effects prevent the breaking of electroweak
symmetry by the Higgs mechanism.  As the universe expands and the temperature
falls, the plasma supercools in this ``symmetric'' phase until the
probability of bubble nucleation is large enough that bubbles of the
lower temperature ``asymmetric'' phase,
in which a Higgs field condensate breaks
electroweak symmetry, form.  These bubbles are thermodynamically favorable,
so they expand, converting the symmetric phase into the asymmetric phase.
At the bubble surfaces, where the phase conversion occurs, the plasma is
thrown out of equilibrium by the motion of the phase boundary, or bubble
wall.  Inside the bubbles, baryon number is approximately conserved; outside
it is rapidly violated. If the departure from equilibrium on the bubble
surface biases the rate of baryon number violation in a CP violating
fashion, this produces a baryon asymmetry.

Specific mechanisms address how the motion of the bubble wall biases the
baryon number violating processes.  The most efficient mechanisms rely on
transport. The bubble wall separates particles and their
antiparticles in a CP violating fashion. In the thin wall case this
occurs through CP violating quantum mechanical reflection
\cite{CKNlepton,CKN,JPTI}, while in the thick wall case the
energy levels of particles and their antiparticles
split in the presence of a $CP$ violating condensate,
leading to a different perturbation in the population densities of particles
and antiparticles
\cite{Zadrozny,JPT,CKN2}. The difference between particle and antiparticle
populations is then
transported into the symmetric phase, where it biases the baryon number
violation rate.  The models based on transport, like most, depend intimately
on the details of the bubble wall shape and its motion. If the bubble wall is
very thin, quantum mechanical reflection is the correct language; if it is
very thick, scatterings are frequent, and the problem should be viewed
semi-classically. Correctly describing the situation where the wall
is thin in comparison to the de Broglie wavelength of
some particles, but thick enough that particles scatter
frequently on the wall, is still an open problem.
In the thin wall case the main effect comes from the
low-momentum particles while in the thick wall case the effect from thermal
particles  dominates. The efficiency of transport strongly depends on the
velocity and thickness of the wall. For instance, for a slow wall,
particles diffuse far in front of the wall, and nonlocal
baryogenesis is possible. For a thin and supersonic wall a typical reflected
particle travels about one diffusion length ahead of the wall,
while for a thick and fast wall
no forward transport is possible. In the latter case only local baryogenesis
occurs, {\it i.e.\/} $CP$ violation and baryon violation take place at the
same point in space.  In any case it is clearly of crucial importance
to know the shape and velocity of the expanding bubble wall.

While there has been substantial progress on this problem
\cite{Neil,DineLinde,Mac,Khleb,Arnoldwall}, much remains to be done to compute
the wall velocity and shape reliably.  In particular, no work fully
addresses the following issues: $(a)$ infrared boson populations;
$(b)$ effects of transport; $(c)$ systematic treatment of all relevant
scattering and decay diagrams to leading order; $(d)$ change in velocity and
temperature of the plasma due to latent heat release; $(e)$ dynamical
determination of the wall shape (thickness).  In a recent letter \cite{paper1}
we developed a set of techniques and approximations which address (b),
(c), and (e).
In this paper we present the details of that work and extend it to
include (d).
The problem of infrared boson populations is of
theoretical interest, since it is related to the infrared properties of
gauge theories at finite temperature,
which are not well understood. We leave the proper treatment
of this problem to a future work. The fluid approximation we use
probably underestimates the friction from infrared bosons, so
our result could be interpreted as an upper bound for the wall velocity.
Also note that we work within the Minimal Standard Model, even though
extensions with more CP violation are generally considered more viable
candidates for baryogenesis.
We do this for simplicity, developing the techniques to control the problem
before we attempt to apply them to more complicated and interesting models.

Now let us briefly outline the paper.
In section II, we derive a semi-classical equation of motion
for the Higgs condensate in
the presence of out of equilibrium particle populations.
The semi-classical approximation is accurate provided the scale
on which the Higgs field condenses, given by the wall thickness,
is large in comparison to the inverse momentum scale of particles which
significantly contribute to friction.  Solving this equation requires
knowledge of the effective potential, the temperature at the wall, and
the departure of particle populations from equilibrium in the presence of
the wall.  In section III we review the state of the art in the effective
potential.  In section IV, we study the bubble nucleation using Langer's
formalism and the potential
of section III. This formalism was applied to the field theory by Coleman and
Callan in \cite{ColemanCallan} and to the electroweak phase transition in
\cite{Enqvist,Neil,DineLinde,Mac,CarrKap} and permits the
calculation of the temperature at which the bubbles nucleate.

The liberation of latent heat produces a jump in temperature and velocity
across the wall, and changes the temperature behind the wall with respect to
the temperature far from the bubble.
We compute this effect using stress-energy
conservation, as done for example in \cite{Enqvist,Mac,Heckler}.
Including this effect could be important since it may
significantly increase the friction on the
wall, especially when the wall velocity approaches the speed of sound.
This is an important improvement on our work in
\cite{paper1}, where for simplicity we ignored this effect.

Next, we  address  the departure of particle populations from equilibrium
in the presence of the wall.  In section V we solve exactly
a particularly simple case of free particle scatterings off the wall (assuming
efficient diffusion so that no piling up occurs in front of the wall).
Unfortunately this is a reasonable approximation only for an unrealistically
thin wall. However, it does help us
to determine what aspects of the departure from equilibrium are most important.
For fermions we find that most of the friction arises from thermal
energy particles, while for bosons most arises from infrared particles.
The friction depends on a particle's mass as $m^4$, so
top quarks appear to be the largest contributors to the friction, even
though they are scattered fairly efficiently.
This justifies the development
of an approximation for the particle populations which models thermal
particles reasonably accurately.  This is the fluid approximation, which
is developed in some detail in section VI, with a systematic study of the
leading order tree-level processes that restore equilibrium postponed to
Appendix A.  Section VII studies aspects of transport in the fluid
equations by finding the response to a $\delta$-function source. It is shown
that, in contrast to a slow wall, no forward transport is possible for a
supersonic wall. The accuracy of and possible
improvements to the fluid approximation within the context of a momentum
expansion  are discussed in Appendix B.

Finally, we combine the fluid approximation with the equation of motion
for the Higgs field and solve for the bubble velocity and profile.
In spite of the nonlinear character in the equation of motion,
this system of equations can be solved analytically using the Fourier
transform and a two parameter {\it Ansatz\/} for the wall shape, as we show in
section VIII.
In section IX we describe a numerical technique which allows a general wall
profile. The results are presented in section X.
The numerical work shows that the {\it Ansatz}  models the wall velocity
with good accuracy (typically within $5\%$) although it is not as good at
modelling the wall profile, as illustrated in Figure \ref{Blah}.
We conclude that the {\it Ansatz} technique
is reasonably accurate for subsonic wall velocities.  However, for
very supersonic velocities, which may occur if infrared physics generates
gauge condensates in the symmetric phase,
both the {\it Ansatz}, and the fluid approximation
itself, break down.  This is discussed in Appendix C, where it is shown
that the bubble wall cannot propagate as an ultrarelativistic detonation.

For the impatient reader we suggest a fast track to reading the paper:
section II, the second half of section IV, section VI, and sections VIII -- X.

\section{Equation of motion}

We are interested in the dynamics of an infrared Higgs condensate,
which we will treat as a classical background field.  From the terms in
the Electroweak Lagrangian containing $\Phi$,
\[
{\cal L} =  ({\cal D} _{\mu} \Phi ) ^{\dag} {\cal D} ^{\mu} \Phi
\! + \mu \Phi^{\dag} \Phi - \lambda (\Phi^{\dag} \Phi)^2 -
\sum y ( \Phi ^{\dag}
\overline{{\psi}_{R}}
\psi _{L}\! +\! \Phi \overline{{\psi}_{L}} \psi_{R})
\]
where the sum is over all massive fermions and $y$ is the Yukawa coupling,
we derive the operator equation of motion
\[
\Box \hat{\Phi} + i g \hat{A}^{\mu} \partial_{\mu} \hat{\Phi}
+ \frac{ig}{2} (\partial \cdot \hat{A}) \hat{\Phi} - \frac{g^2}{4}
\hat{A}^2 \hat{\Phi} - \mu \hat{\Phi}
 + 2 \lambda (\hat{\Phi}^{\dag} \hat{\Phi} )
\hat{\Phi} + \sum y \hat{\overline{\psi_R}} \hat{\psi}_L
\]
We have suppressed group  indices.
Each term containing $A$ actually appears once for weak isospin and once for
hypercharge.

This operator should annihilate the physical thermal state.
We will shift $\Phi$ by a classical part,
$\Phi = \Phi_{cl} + \delta \Phi$, choosing $\delta \Phi$ such that
$\langle \delta \Phi\rangle = 0$.  We then evaluate the trace of the
operator equation of motion over
the (out of equilibrium) thermal density matrix describing the propagating
bubble wall.  We will assume that there are no charge conjugation violating
gauge condensates, so that $\langle A^{\mu}\rangle = 0$.
We then find
\[
\Box \Phi_{cl}  - \mu \Phi_{cl} +
 2 \lambda (\Phi^{\dag}_{cl} \Phi_{cl}) \Phi_{cl} +
\]
\[
2 \lambda \left( 2 \langle\delta \Phi^{\dag} \delta \Phi\rangle \Phi_{cl} +
 \langle\delta \Phi^2\rangle \Phi^{\dag}_{cl} \right)
- \frac{g^2}{4} \langle A^2\rangle +
\sum y \langle \overline{\psi}_R \psi_L\rangle = 0
\]
We assume that $\Phi_{cl} = [ 0 \; \; \phi/\sqrt{2}\; ]^T$ (which is the
same as neglecting charge conjugation violating condensates).
Next, we evaluate the thermally averaged operators using WKB wave functions.
This makes sense because the background field $\phi$ varies on a scale
much longer than $T^{-1}$, which characterizes the reciprocal momenta of
particles in the plasma.  In this approximation,
\[
\langle A^2\rangle = \langle A_{vac}^2\rangle +
\sum \int \frac{d^3 k}{(2\pi)^3 E} f(k,x)
\]
where the sum is over group  indices  and spins and $f$
is the phase space population density, and
\[
\langle \overline{\psi_R} \psi_L\rangle =
\frac {1}{2}\langle \overline{\psi} \psi\rangle _{vac} + \sum
\int \frac{d^3k}{(2\pi)^3} \frac{m}{2 E} f(k,x)
\]
where the sum is over group and Dirac  indices.

The equation of motion becomes
\begin{equation}
 \Box \phi + V'(\phi ) + \sum {dm^2\over d\phi}
\int \frac{d^{3} k}{ (2 \pi )^{3}\; 2E}  f(k,x) = 0
\label{eq:equation for phi ii}
\end{equation}
where $V$ is the renormalized vacuum
potential, and the sum includes all massive physical degrees of
freedom, including the high frequency parts of the Higgs field.
Note the condensed notation: $m^2=y^2\phi^2/2$ for quarks
and leptons, $g_w^2 \phi^2/4$ for the gauge fields,
$ 3 \lambda \phi^2 - \mu$+(thermal) for
the Higgs bosons, and $\lambda \phi^2 - \mu +$(thermal)
for the pseudo-Goldstone modes.  This equation has also been derived by
diagrammatic techniques in \cite{Mac}.

This equation has a nice physical interpretation.
Multiplying it by $\partial^\mu \phi$, we find
\begin{equation}
0 = \Box \phi \partial^\mu \phi + \partial^{\mu} V + \int \frac{d^3 k}{
(2\pi)^3 2E} f(k) \partial^{\mu} m^2
\label{Newtonforphi}
\end{equation}
The first term can be recast as
 $\partial_\mu\partial^\mu\phi\partial^\nu\phi=
\partial_\mu(\partial^\mu\phi
\partial^\nu\phi)-\partial^\nu(\partial_\mu\phi\partial^\mu\phi/2)$
, so that when taken together
the first two terms are the divergence
of the stress-energy of the Higgs field
$T^{\mu\nu}(\phi)=\partial^\mu\phi\partial^\nu\phi-g^{\mu\nu}{\cal L}(\phi)$,
where ${\cal L}(\phi)=\partial^\mu\phi\partial_\mu\phi/2-V(\phi)$
is the Higgs lagrangian.
The last term represents exchange of stress-energy with particles;
it looks like $\int d^3k f(k)  (-F^\mu) /(2 \pi)^3$.
Here $F^\mu=-\partial^\mu E=(-\partial_t E, -\vec \nabla E)$ is the 4-force a
particle feels in the presence of the wall; the wall feels an equal and
opposite force.
Thus we can recast (\ref{Newtonforphi}) as follows
\begin{equation}
\partial_\nu T^{\nu\mu}(\phi)-
\int \frac{d^3 k}{
(2\pi)^3 } f(k) F^\mu =0\,,\qquad F^\mu=-\partial^\mu E
\label{Newtonforphii}
\end{equation}
The equation of motion for the Higgs field in the WKB approximation
splits the total (conserved) stress-energy tensor
$T^{\mu\nu}$ into Higgs and fluid parts
({\it sub-systems}) in a natural way,
and provides the prescription for how each is violated.

As an application of (\ref{Newtonforphii}),
$\int ({\rm Eq.} \ref{eq:equation for phi ii}) \phi' dz$ is
the total pressure on the wall.  If this integral does not vanish, then the
wall accumulates momentum and accelerates; hence this integral must vanish
for a steady state wall.  This fact forms the basis of the analysis
of \cite{Neil,DineLinde,Mac}.

We write the population density $f$  as the equilibrium
population $f_0$ plus a deviation, $f = f_0 + \delta f$.
The vacuum contribution $V'(\phi)$ and
the contribution from $f_0$ combine to give the finite
temperature effective potential $V_T ' (\phi)$.  Thus we have

\begin {equation}
\label{eqmo}
 \Box \phi + V_{T}'(\phi) + \sum { dm^2\over d\phi} \int
\frac{d^{3}p}{(2 \pi )^{3}
\, 2 E}  \delta f(p,x)  =0
\end{equation}
If the system were in phase equilibrium we would have expected
$\Box \phi + V_T'$;
we see that the frictive force (dissipation) arises due to the departure from
thermal equilibrium $\delta f$ (fluctuation).
Our goal will be to solve this equation for a stationary wall, {\it i.e.\/}
after the wall has reached a steady shape and velocity.  To do so we need to
know the effective potential $V_T$ and the temperature $T$, and we need a way
to calculate $\delta f$.

\section{Effective Potential}

The high temperature expansion of the one loop effective
potential, ignoring scalar loops, is \cite{DineLinde}
\begin{equation}
V(\phi,T) = D (T^2 - T_o^2) \phi^2 - E T \phi^3 +
{\lambda_T\over 4} \phi^4 \ .
\end{equation}
Here
\[
D = {1\over 8v_o^2} ( 2 m_W^2 +
m_Z^2 + 2 m_t^2) \sim 0.169 \ ,
\]
\[
E =  {1\over 4\pi v_o^3} ( 2 m_W^3 +
m_Z^3) \sim 10^{-2} \ ,
\]
\[
T^2_o = {1\over 4D}(m_H^2 - 8Bv_o^2) \ ,
\]
\begin{equation}
\lambda_T = \frac{m_H^2}{2 v_o^2} - {3\over 16 \pi^2 v_o^4}
\left( 2 m_W^4 \ln{m^2_W\over a_B T^2} +
m_Z^4 \ln{m^2_Z\over a_B T^2} -
4 m_t^4 \ln{m^2_t\over a_F T^2}\right) \ ,
\label{lambdaTeq}
\end{equation}
where $\ln a_B = 2 \ln 4\pi - 2\gamma \simeq 3.91$,
$\ln  a_F = 2 \ln \pi - 2\gamma \simeq 1.14$, and
\[
B = {3\over 64 \pi^2 v_o^4} (2 m_W^4 +
m_Z^4 - 4 m_t^4)
\]

The value of $\phi$ in the broken phase at equilibrium is easily evaluated
to be $2ET/\lambda_T$.  If we take $\lambda_T $
parametrically $\sim g^2$, which
is natural from the renormalization structure of the standard model, we
find $\phi \sim gT$, which is small enough that the perturbation series
may not be well behaved \cite{Linde}.  We should therefore include the
two loop contribution, which has recently been computed by \cite{Arnold}
and \cite{Fodor}.  The two loop expression contains terms $\sim g^4 \phi^2
T^2$, with a coefficient including constants and logs of the temperature
over the renormalization scale, which slightly corrects $D$.  We will drop
this term as it has no influence on the behavior of the phase transition.
Two more important corrections do occur, however.  One is that the
longitudinal components of the gauge bosons acquire large plasma masses
\cite{Carrington,DineLinde} and do not contribute to the strength
of the transition.  Following \cite{DineLinde} we drop them from
the $E$ term in the effective potential, giving
\[
E = \frac{1}{12\pi}(4 m_W^3 + 2 m_z^3 + (3 + 3^{1.5}) \lambda_T^{1.5})
\]
 where the
$\lambda_T$ dependent term comes from including ring improved scalar loops.
The other correction is that qualitatively new terms
$ \sim g^4 \phi^2 T^2 \ln(m/T)$ appear.  Here $m$ stands for sums of masses
of particles.  In the approximation that the Higgs masses are small in the
symmetric phase (which should be good near the transition temperature) we
may treat $m \sim g \phi$; the log of $g$ is absorbed into $D$, and the
new term becomes $-C \phi^2 T^2 \ln(\phi/T)$.
The value of the coefficient is \cite{Fodor}
\[
 C \simeq \frac{1}{16\pi^2} (1.42g_{w}^{4}
+ 4.8 g_{w}^{2} \lambda_T - 6 \lambda_T^{2})
\]
 where we
have again left out contributions from transverse gauge bosons.  Including
this term, the value of $\phi$ in the broken phase at equilibrium
becomes $\phi/T = E/ \lambda_T + \sqrt{ (E/\lambda_T)^2 + 2C/ \lambda_T}$,
which indicates that the new term strengthens the transition and contributes
at the same parametric order as the one loop term $E$, although for the
parameters we will be interested in its contribution is smaller.

Because the two loop result is of the same parametric order as the one loop
result we might worry that perturbation theory cannot establish the strength
of the transition reliably.  In fact we expect that the perturbative
computation of the effective potential should break down in the symmetric
phase.  This means that the value of $V_T(0)$ may be shifted somewhat from
zero.  Shaposhnikov has suggested that such a shift may arise due to
the formation of gauge condensates in the symmetric phase \cite{Shapnonpert}.
He advocates adding a term $ - A_F g^6/12 * {\rm Pit}(\phi)$ to the
effective potential to account for such an effect.  Here the function
Pit describes the $\phi$ dependence of the strength of such a condensate,
and is smooth at $\phi = 0$ but falls as $\exp(-\phi/g^2)$ at larger $\phi$.
We will include such a contribution to parametrize our ignorance of the
effective potential in the symmetric phase.  We choose Pit=sech$(
\phi/\alpha_W)$ because it is simple, but our results do not depend
strongly on the functional form or the exact rate of exponential decay,
as long as that decay is rapid.  We treat the value of $A_F$
as an unknown parameter; there is some lattice evidence for its
value but it is still preliminary \cite{Kajantie,Fodor2}.

Our final form for the effective potential is then
\begin{equation}
V_T(\phi) = D (T^2 - T_0^2) \phi^2 - C T^2 \phi^2 \ln(\frac{\phi}{T})
           - E T \phi^3 + \frac{\lambda_T}{4} \phi^4
	  - \frac{A_F g^6}{12} {\rm sech}(\frac{\phi}{\alpha_W})
\label{potential}
\end{equation}
\section{Nucleation and Hydrodynamics}
\label{hydrosec}

We must next determine what $T$ is at the bubble wall.  Following
\cite{Neil,Enqvist} we do this in two steps; first we calculate the
temperature in the universe when most bubble nucleation events occur,
and then we calculate the temperature correction due to the release of
latent heat by the bubble as it expands.

The process of bubble nucleation at finite temperature was worked out by
Linde \cite{Lindenuc} and applied to the electroweak phase transition in
\cite{Enqvist,Neil,DineLinde,AndersonHall,Mac,CarrKap}.  The basic idea
is that small bubbles are thermodynamically unfavorable because of positive
surface free energies; to become large enough not to re-contract a bubble
must pass over a free energy barrier.  The lowest route over the barrier goes
through a saddlepoint configuration of the effective action called the
critical bubble.  It is a spherically symmetric solution to the classical
equilibrium equation of motion
\begin{equation}
\nabla ^2 \phi (r) = V_T '(\phi) \qquad
\phi'(r=0)=0\: , \: \phi(\infty) = 0
\label{critbubble}
\end{equation}

 The free energy of such a solution is
\begin{equation}
S_{\rm crit} = \int d^3 x \left( \frac{1}{2}
(\nabla \phi)^2 + V_T(\phi) \right)
\end{equation}
There are analytic formulas for $S_{\rm crit}$ only in special cases; in
general it must be determined numerically.

The nucleation rate per unit volume per unit time is
\begin{equation}
\frac{\Gamma}{ V}  =  I_0 T^4 \exp(-\frac{S_{\rm crit}}{T})
\end{equation}
Carrington and Kapusta have recently computed the prefactor \cite{CarrKap}.
Numerically evaluating the expressions in their paper we find, roughly,
$\ln(I_0) \simeq -14$.  The exact value depends on the parameters of the
effective potential and (weakly) on temperature, but as we will see we
only need a rough estimate.

We can now compute the action of the critical bubble at the time when
most of the universe changes phase by following the technique of
\cite{AndersonHall}.
If bubbles expand at a velocity $v_w$ which depends only weakly on temperature,
then the fraction of space remaining in the symmetric phase at time $t$ is
roughly
\begin{equation}
\exp \left( - \int_{- \infty }^{t} \frac{4 \pi}{3} v_w^3 (t-t')^3
     I_0 e^{-S(t')/T} dt' \right)
\label{bubble overlap}
\end{equation}
where the exponential accounts for bubble overlap \cite{someone}.
Expanding the bubble action about the point when the above integral is 1,
$S = S(T_{\rm nuc}) + (t-t') dS/dt$ and using $d/dt = (dT/dt)d/dT$ and
$ (1/T) dT/dt = H$ the Hubble constant, we find that most of space has
changed phase when
\[
e^{S(T_{\rm nuc})/T} = \frac{8 \pi v_w^3 I_0}{ (H T dS/dT)^4}
\]
A reasonable estimate for $T dS/dT$ is the thin wall value
$2ST/(T_{\rm eq}-T) \sim 10^4$, which roughly agrees with the value we
find by numerically evaluating $S/T$ at closely spaced values of $T$.  The
value of $H$ follows from the Friedmann equation
\[
H^2 = \frac{8 \pi G}{3}\: \frac{ \pi^2 g_* T^4}{30}
\]
where the second factor is the energy density of the plasma and $g_* =
106.75 (1 + O(\alpha_s))$ in the symmetric phase (the $O(\alpha_s)$ arises
from thermal masses and other interaction effects).  The value
of $v_w$ is to be determined, but since a subsonic bubble is proceeded by
a hydrodynamic front travelling at the speed of sound which raises the
temperature of the fluid and prevents  nucleation,  it is sensible to
choose the speed of sound $v_s = 1/\sqrt3$.

Putting all the expressions together and
using $T \sim 100$GeV, we find $S/T \simeq 97$.
In Ref. \cite{AndersonHall} a slightly higher value was
found because of the value for $I_0$ we use.   We may
now determine the temperature at which most  nucleation occurs   by
solving (\ref{critbubble}) for various values of $T$ until we find one which
gives $S_{\rm crit} = 97$.

This gives the temperature around the bubble at the time that it nucleates,
but we are really interested in the temperature at the bubble wall while
it is expanding.  This will be elevated from $T_{\rm nuc}$ because of the
liberation of latent heat as the symmetric phase
is converted into the asymmetric phase.

In \cite{Neil,DineLinde,Mac} it is argued that the elevation of temperature
at the bubble wall is not important in determining its velocity.
This is
true when the change in temperature, which is $\sim l/(d\rho /dT)$ (where
$l$ is the latent heat), is small compared to the supercooling, $T_{\rm eq}
-T_{\rm nuc}$.  \cite{Neil,Mac} show explicitly that for the effective
potential they consider, $  (T_{\rm eq} -T_{\rm nuc})(d\rho/dT)$ is about
5 times $l$.  However, that analysis is based on a small assumed value of
$m_t$; the latent heat depends on $D \propto m_t^2$, and the supercooling
$T_{\rm eq}-T_o \propto 1/D$; so for $m_t \simeq 174$GeV,
the temperature increase probably is important.

To compute the temperature at the wall we need to solve the temperature
profile around an expanding bubble wall.
The equation of state in the
symmetric phase is $\rho_s = (\pi^2 g_* / 30)T^4, \; p_s = \rho_s/3$, and in
the broken phase it is $\rho_a = \rho_s + l(T), \; p_a = p_s -V_T(\phi(T))$.
Here $l$ is the latent heat and is given by
\begin{equation}
l = V_T(\phi_a) - T \frac{dV_T(\phi_a)}{dT} = - D \phi^2 (T^2 + T_o^2)
+ (\frac{\lambda_T}{4} - B) \phi^4 \simeq -2D \phi^2 T^2
\end{equation}
Note that $l/\rho \sim 0.01 <<1$, and similarly we find $(T_{\rm eq} - T)/T
\sim 0.01 <<1$, so it is reasonable to expand to lowest order in these
 quantities.   However, $l$ turns out to be quite a strong function of
temperature,
\begin{equation}
T \frac{d(l/T^4)}{dT} \simeq - 4D \frac{\phi}{T} \frac{T \, d(\phi/T)}{dT}
\end{equation}
We can determine $T d(\phi/T)/dT$ in the broken phase as follows; the value
of $\phi$ is determined by $V_T'(\phi)=0$, and since nonperturbative effects
are tiny in the broken phase we can use the two loop expression, which gives
\begin{eqnarray}
0 & = & 2D(T^2 - T_o^2) \phi - C \phi^2 T^2 (2 \ln\frac{\phi}{T}+1) -
3E\phi^2 T +
\lambda_T \phi^3  \nonumber \\
 0 & = & 2D(1-\frac{T_o^2}{T^2}) - 2C\ln\frac{\phi}{T} -C -3E \frac{\phi}{T}
+\lambda_T \left (\frac{\phi}{T}\right )^2
\end{eqnarray}
Since this equals zero, its total derivative with respect to $T$,
 $\partial/\partial T + d(\phi/T)/dT$ $ \partial/\partial(\phi/T)$,
should vanish.  This gives
\begin{equation}
\frac{d(\phi/T)}{dT} = \frac{-4D T_o^2 \phi /T}
{2 \lambda_T \phi^2 - 3 E \phi T - 2 C T^2}
\end{equation}
and hence
\begin{equation}
T\frac{d(l/T^4)}{dT} \simeq -\frac{4D\phi}{T} \frac{-4D T_o^2 \phi /T}
{2 \lambda_T \phi^2 - 3 E \phi T - 2 C T^2}
\end{equation}
For typical values of $\lambda_T = 0.03 , \phi = 0.8$, we find $Td(l/T^4)/dT
\simeq 18$,
which is about half as large as $\rho/T^4$.
This rather surprising result occurs
because $\phi$ is strongly temperature dependent in the
broken phase near the transition, and because the top quark is so massive;
around the transition temperature the number of top quarks is rapidly freezing
out.  Because $\rho$ turns out to have this extra temperature dependence,
the speed of sound in the broken phase, $ dp/d\rho = (dp/dT)/(d\rho/dT)$,
is around $15 \%$ smaller than in the symmetric phase.  Thus it is reasonable
to expand to linear order in $l/\rho$ and $(T-T_{eq})/T$,
and hence in $\delta T$ and $v$, but not to neglect $dl/dT$.

Now let us explore the fluid temperature and velocity in the vicinity of an
isolated spherical bubble.  From Eq. (\ref{bubble overlap}) and the argument
which follows, we find that the typical bubble grows to a radius of
$\sim 1/(HT dS/dT) \sim 10^{12}/T$, while the distance it takes for a bubble
to accelerate to a steady velocity is around the ratio of the surface
tension to the free energy density difference of the two phases, which is
of order $50/T$.  On the scale of the bubble radius, then,
we can treat the plasma as a perfect
fluid and expect the temperature and velocity
to be functions of $r/t \equiv x$ only, where time $t$ is measured from
the time of nucleation.
Conservation of the stress energy tensor gives the equations
\begin{equation}
0 = \partial_{\mu} T^{\mu \nu} = \partial_{\mu} ( (\rho + p) u^{\mu} u^{\nu} -
p g^{\mu \nu}) = u^{\nu} u \cdot \partial (\rho + p) + (\rho + p) (u^{\nu}
\partial \cdot u + u \cdot \partial u^{\nu}) - \partial^{\nu} p
\end{equation}
which to linear order in $v$ and $\delta T$ (which are linear in $l/\rho$) give
\begin{equation}
(\rho + p)\left (\frac{2}{x} v+ \frac{dv}{dx}\right ) - x \frac{d\rho}{dT}\,
\frac{dT}{dx} = 0 \qquad
 \frac{dp}{dT}\frac{dT}{dx} = (\rho + p) x \frac{dv}{dx}
\label{hydro}
\end{equation}
The $2v/x$ arises from taking a divergence in spherical coordinates and is
the only difference between the spherical geometry and the planar geometry
studied by \cite{Enqvist,Mac}.

Because the equation of state for the symmetric phase is simple we can
solve the above equations for the case that the wall velocity is less than
the speed of sound in the broken phase (in which case $T$ is constant and $v$
is zero inside the bubble).  The equations become
\begin{equation}
x \frac{dv}{dx} = \frac{d \ln T}{dx} \qquad 2v +
x\left (1-\frac {x^2}{v_s^2}\right ) \frac{dv}{dx} = 0
\label{gasseq}
\end{equation}
which give
\begin{equation}
v = A \left (1+\frac{v_s}{x}\right )\left (1-\frac{v_s}{x}\right )
\qquad \ln \frac{T}{T_{\rm nuc}} \simeq \frac{T - T_{\rm nuc}}
{T_{\rm nuc}} = 2 A v_s \left (1 - \frac{v_s}{x}\right )
\end{equation}
Here $A$ is an undetermined parameter and $v_s = 1/\sqrt{3}$.  It is
interesting to note that, in this approximation (linear order in $l/\rho$), the
temperature and velocity are actually continuous at $x=v_s$, although the
first derivative is discontinuous.
This holds to all perturbative orders in $l/\rho$.  The shock front is
exponentially weak, as found in \cite{Heckler}.

Since the wall does not accumulate energy or momentum,
Eq. (\ref{hydro}) also provides us with boundary conditions at the bubble
wall, by replacing the derivatives with differences.  Because the wall
is thin compared to the bubble radius the term $2v/x$ is negligible and we
get
\begin{equation}
(\rho_s + p_s) v - v_w \rho_s = -v_w \rho_a(T_{\rm inside})
 \qquad -v_w (p_s + \rho_s) v + p_s = p_a(T_{\rm inside})
\label{boundary condition}
\end{equation}
which, at fixed $v_w$, allows us to solve for the temperature in front of
and behind the wall.  Defining $y = v_w/v_s$, we find
\begin{equation}
\frac{T_{s}-T_{\rm nuc}}{T_{\rm nuc}} = \frac{y^2 l(T_{a})}{2(1+y)^2(1-y)}
\qquad
\frac{T_{a}-T_{\rm nuc}}{T_{\rm nuc}} = \frac{y^2 l(T_{a})}{4(1+y)^2}
\end{equation}
Note the dependence of $l$ on $T_{a}$ the temperature inside the bubble,
which makes these equations  transcendental.

When the wall velocity exceeds $v_s$, the only causal solution to
(\ref{gasseq}) outside the bubble is $v=0,\:T=T_{\rm nuc}$.  We may find
the temperature just inside the bubble wall directly from the boundary
conditions
\begin{equation}
(\rho_a + p_a)v - v_w \rho_a = -v_w \rho_s(T_{\rm nuc}) \qquad
-v_w(\rho_a + p_a)v + p_a = p_s(T_{\rm nuc})
\end{equation}
which give
\begin{equation}
\frac{T_{a}-T_{\rm nuc}}{T_{\rm nuc}} = \frac{y^2 l(T_{a})}{4(y+1)(y-1)}
\end{equation}
In both cases the temperature is elevated on one or both sides of the bubble.
The elevation has a $1/(y-1)$ behavior, which diverges near $v_s$.  This
divergence means that our expansion in small $v$ and $T-T_{\rm nuc}$ must
break down very near $v_s$; but there is still a very large elevation in
temperature when $v \simeq v_s$.  As the symmetric phase becomes more favorable
at higher temperatures this prevents the wall from propagating close to the
speed of sound; it must either be a deflagration or a strong detonation.

We have not treated the case where $v_w$ lies between $v_s$ and the speed
of sound in the broken phase; this case is much more complicated because
the fluid velocity is nonzero on both sides of the wall and we need three
conditions to determine all unknowns.  Fortunately, in this case the
temperature would be elevated enough that the wall probably cannot propagate
at this speed; so we will not treat this case.
We will also not treat the
effect of latent heat released by other bubbles, although as the phase
transition progresses and bubbles begin to collide, one bubble's wall will
frequently be in the hydrodynamic wave initiated by another bubble.  These
effects make the later phases of the transition quite complicated, and
the wall velocity we compute here should be an upper bound on the velocity
in this epoch \cite{Heckler}.
This effect should be important especially
when the wall velocity is much smaller than the speed of sound. In this case
bubbles would expand most of the time in the background of a sound wave of
another bubble. This effect may significantly enhance baryon production as
argued in \cite{Heckler}.

\section{Free Particle Approximation}

Let us proceed to computing the wall velocity.  First we will discuss a
simple limit, the case where mean free paths are enormously longer than the
wall.  This limit has been treated thoroughly by \cite{DineLinde,Mac} and
we will only discuss it briefly to illuminate a few points about which
particles provide most of the friction.

In the WKB (semi-classical) approximation particles should obey a Boltzmann
equation,
\begin{equation}
\label{Boltzmann}
\partial_t f + \dot{\vec{x}} \cdot  \partial_{\vec{x}}f
+ \dot{\vec{p}} \cdot \partial_{\vec{p}} f = -C[f]
\end{equation}
Here $\dot{\vec{p}} = -\hat{z} \partial E / \partial z = (1/2E)dm^2/dx$,
$\partial_t f = +v_w \partial_z f$ (where we choose the broken phase at
$z=+\infty$ so the wall moves to the left)
and $C[f]$ is the collision integral.  The free particle approximation
consists of dropping $C[f]$.  Since collisions are what drive $f$ to
equilibrium, we must insert equilibrium by hand as a boundary condition for
incoming particles.  We will use this approximation (though we should note
that some of the departure from equilibrium may reflect back onto the wall
\cite{Arnoldwall}).
In this case the Boltzmann equations can be solved
exactly,
\begin{eqnarray}
f^{-1} = \exp \left( \frac{E + v_w p_z}{(1-v_w^2)T}
- \frac{\gamma v_w}{T} \left(
\frac{(p_z + v_w E)^2}{1-v_w^2} + m^2 \right)^{\frac{1}{2}} \right) \pm 1\,,
\; p_z > -\sqrt{m_0^2 - m^2}
\nonumber \\
= \exp \left( \frac{E + v_w p_z}{(1-v_w^2)T} + \frac{\gamma v_w}{T} \left(
\frac{(p_z + v_w E)^2}{1-v_w^2} + m^2 \! - \! m_0^2 \right)^{\frac{1}{2}}
 \right)  \pm 1\,,
\;  p_z < -\sqrt{m_0^2 - m^2}
\end{eqnarray}
Here $m_0$ is the mass in the broken phase and $\pm$ is $+$ for fermions
and $-$ for bosons.  Determining $\int d^3k/((2\pi)^3 2E) f$, the term
in Eq. (\ref{eqmo}), for this
expression appears hopeless, but we can expand to lowest order in $v_w$,
which gives
\begin{eqnarray}
& &  \int  \frac{d^3p}{(2\pi)^3 2E} f_0 +  \nonumber \\
& &  v_w  \int  \frac{d^3p}{(2\pi)^3 2E} \frac{e^{E/T}}{(e^{E/T} \pm 1)^2}
\left\{ \begin{array}{ll}
\sqrt{p_z^2 + m^2 } - p_z\,, & p_z > - \sqrt{m_0^2 - m^2} \\
-\sqrt{p_z^2 + m^2 - m_0^2} - p_z\,, & p_z < - \sqrt{m_o^2 - m^2} \end{array}
\right.
\end{eqnarray}
This term is independent of the momentum
perpendicular to the wall, $p_\bot$; so we can perform the integral over
$p_\bot$.
\[
\int \frac{d^2p_\bot}{(2\pi)^2 2E} \frac{e^{E/T}}{(e^{E/T} \pm 1)^2} =
\frac{f_0(\sqrt{p_z^2 + m^2})}{4 \pi}
\]
Integrating this expression times $mm'$ over the wall yields the velocity
dependent part of the pressure; the result agrees with the integral found
by \cite{DineLinde}, who have evaluated it to order $m^4$.

Let us examine the integral over $p_\bot$ more carefully.  For bosons,
the integral is
\[
\int \frac{p_\bot dp_\bot}{4 \pi E}\: \frac{1}{4} {\rm csch}^2\frac{E}{2T}
\]

At small $E$, ${\rm csch}^2(E/2T) \simeq 4 T^2/E^2$ and the integral has
a linear small momentum divergence, cut off by $\sqrt{p_z^2 + m^2}$; the
dominant contribution is infrared.  The resulting integral over $p_z$ is
also linearly IR divergent, as both $f$ and
$ \sqrt{p_z^2 + m^2} - p_z$
behave as $1/E$ above $m_0$.  Thus, the friction from $W$ bosons arises
primarily from very infrared ($E \sim m$) particles.
This is because the Bose
distribution function has a pole at $E=0$, and because a particle's
contribution to the Higgs equation of motion goes as $1/E$.

Fermions have much more mild infrared behavior; the
integral over transverse momenta is
\[
\int \frac{p_\bot dp_\bot}{4 \pi E}\: \frac{1}{4} {\rm sech}^2\frac{E}{2T}
\]
and at small $E$ the function ${\rm sech}^2(E/2T)$ behaves as 1; the integral
is well behaved in the infrared, even for $\sqrt{p_z^2 + m^2}=0$, and
the dominant contribution is from particles with $p_\bot \sim (1-2) T$.
However, the integral over $p_z$ {\it is} small $p_z$ divergent, though only
 logarithmically;  so the dominant contribution from top quarks comes from
particles with small transverse momenta, but thermal energies, i.e. from
particles at glancing incidence.  This is a combined effect of the large
phase space in such particles and their long time on the wall.

Now let us examine the analysis of Dine et. al. \cite{DineLinde}.  They
propose to treat the particles in a relaxation time approximation.
However, they compute the particle population at a point
as having come off of a sheet a fixed $z$ away from the point of interest.
A particle's lifetime then actually goes as
$ \tau p/p_z = \tau \sqrt{1 + (p_\bot /
p_z)^2}$, so the relaxation time is longer for particles
travelling at glancing incidence.  Their result then contains a logarithmic
enhancement, which can be traced to these glancing incidence particles.

We will now show that no such log
occurs in a relaxation time approximation when the
relaxation time is short compared to the thickness of the wall.  Writing
$f = f_0 + \delta f$ and working to lowest order in $\delta f$, and treating
the wall as stationary so $\delta f = \delta f(z + v_w t)$,
the Boltzmann equations become
\begin{equation}
(v_w + \frac{p_z}{E}) \delta f' + C[f] = v_w \frac{(m^2)'}{2E} \frac{\exp(E/T)}
{(\exp(E/T) \pm 1)^2}
\label{relaxation}
\end{equation}
The relaxation time approximation is $C[f] = \delta f/\tau$, with $\tau$
independent of momentum.  If $\tau << L$
(with $L$ the wall thickness)
then to lowest order in $\tau/L$ the derivative term may be dropped, giving
\begin{equation}
\delta f = \tau v_w  \frac{(m^2)'}{2E} \frac{\exp(E/T)}
{(\exp(E/T) \pm 1)^2}
\end{equation}

The friction on the wall from one species
in the relaxation time approximation is then
\begin{equation}
\int^{\infty}_{-\infty} dz \phi' \frac{dm^2}{d\phi} \int \frac{d^3k}{(2\pi)^3
2E} \tau v_w \frac{(m^2)'}{2E} \frac{\exp(E/T)}{(\exp(E/T)\pm 1)^2}
\end{equation}
which, at lowest order in an expansion in $m$, integrates to
\begin{equation}
\frac{\tau v_w}{4 \pi^2} \int m^2(m')^2 dz \, , \quad {\rm (fermions)} \qquad
\frac{\tau v_w}{8 \pi} \int m (m')^2 dz \, , \quad {\rm (bosons)}
\end{equation}
which for a tanh wall shape give $\tau v_w m_0^4/40 \pi^2 L$ and
$\tau v_w m_0^3/48\pi L$ respectively.  The top quark contribution is
isotropic, dominated by thermal particles, and has no  log.
The log will be recovered when $\tau >> L$ and the derivative terms
in Eq. (\ref{relaxation}) become important.

We do not expect this approximation to be very good; in particular we should
include the derivative terms and model $C[f]$ more accurately.  This is the
goal of the remainder of the paper.

\section{Fluid Approximation}

In this section we will present a method for calculating
the deviation from equilibrium population density $\delta f$ in (\ref{eqmo})
in the presence of a moving wall.  Our starting point is the
Boltzmann equation,
\begin{equation}
d_t f\equiv\partial_t f +{\dot z}\partial_z f+{\dot p_z}\partial_{p_z}
f =-C[f]
\label{eq:boltzmann}
\end{equation}
where $C[f]$ represents the scattering integral,
$E=(p^2+m^2)^{1/2}$ is
the particle energy, and
\begin{equation}
\dot {z}=v_z={\partial\over \partial {p_z}} E= {p_z\over E}\, , \qquad
\dot p_z=-{\partial\over \partial z} E= -{(m^2)^\prime\over 2E}
\label{eq:velocity acceleration}
\end{equation}
are  the velocity
and  the force on the particle, respectively.
The Boltzmann equation is the semi-classical approximation
to  the quantum Louville equation.  To be valid the background field must
vary slowly enough that particles satisfy the $WKB$ condition:
\[
p>> {1\over L_w}
\]
Since we expect rather thick wall, $L_w>> 1/T$, this relation is
satisfied in abundance for all but the most infrared particles.
We will assume that very infrared $W$ bosons are scattered
up to energies $O(gT)$ quite readily, so $W$ bosons with $E \sim gT$ or larger
are responsible for more friction than very infrared $W$ bosons.
We will use this assumption again and again in what follows.  It may be
wrong\footnote{The analysis of \cite{Mac}, which neglects transport but
attempts to treat the energy dependence of $\delta f$ more carefully,
apparently supports our assumption, but their analysis neglects
thermal corrections to the infrared boson propagator.}
, in which case we need to extend our analysis with a special treatment
of very infrared bosons.  This extension must model the dynamics of
the strongly coupled infrared sector of the thermal field theory, which we
are not able to do at this time.  We will also neglect any friction from the
condensate responsible for $A_F$ (if such a condensate exists) for similar
reasons.  It is our belief that this problem and the problem of determining
the friction arising from thermal and $O(gT)$ particles may be treated
independently, so our analysis may be simply extended once the infrared sector
is understood.

The Boltzmann treatment also requires that scatterings with the plasma are
not too frequent so that particles can be to a good approximation considered
on-shell for all times. To quantify this we write
the Heisenberg energy-time  relation
$\Delta E\sim  \gamma$
where $\gamma$ is the appropriate damping rate which is of order $g^2 T$
\cite{Pisdamprate}.
Since we can only treat particles with $E, p\geq gT$ this
uncertainty can be  ignored.

The Boltzmann equations are nonlinear partial
integro-differential relations and  as such are analytically intractable.
To evade this difficulty we will model each distribution function with a
several parameter {\it Ansatz} called the fluid approximation.  The Boltzmann
equations will then yield a set of ordinary differential equations for
the parameters which will be  tractable  analytically.  The {\it Ansatz} is
$f^{-1} = \exp(E/T - E \delta T/T^2 - \mu/T -p_z v/T) \pm 1$, the form
for a perfect fluid.
We should use different parameters for each species in the plasma, but we
will make an additional approximation that all light degrees of freedom (that
is, all but top quarks, $W$ bosons, and Higgs bosons) are in equilibrium at
a common (space dependent) temperature $T + \delta T_{bg}$ and velocity
$v_{bg}$. We will also give tops and  anti-tops   of both helicity
the same distribution function.  In the Minimal Standard
Model this is reasonable as there is almost no CP violation, and
the difference in transport properties arises only at the
subleading level of weak scatterings.
In two doublet models we might need to be more careful.
Similarly, we will treat $W$ bosons and $Z$ bosons as a single species
(henceforth ``$W$ bosons'') whose mass squared is a weighted average of
the mass squared for the $W$ and the $Z$.  This ensures that the
effective potential parameter $D$ is correct.

For the heavy degrees of freedom we take the distribution function to be
\begin{equation}
f=f_0(E+\delta)={1\over  {\rm e}^{(E+\delta)/T}\pm 1}\,,\quad
\delta= - \left[ \mu + \mu_{bg}+ \frac{E}{T}(
 \delta T+ \delta T_{bg})  + p_z  (v + v_{bg}) \right] \;  \;
\end{equation}
We track three perturbations with respect to the background;
chemical potential $\mu$, temperature $\delta T$  and velocity
 $v$.  This {\it Ansatz} is a truncation of an expansion
in powers of momentum; it gives a reasonable description
for the thermal particles' populations when the  wall background
varies slowly on the scale of the diffusion length.
For top quarks this
should be sufficient as the diffusion length is short and the
influence of infrared particles is phase space suppressed.  We discuss its
validity at some length in Appendix B: suffice it to say that it appears
quite reasonable for top quarks, and quite naive for $W$ bosons.

We now outline the main steps in the derivation of the fluid equations.
First we expand $d_t f$ to linear order in perturbations
\begin{eqnarray}
T d_t f & = & \left[ \dot E+ d_t \delta \right] f_0 ^\prime \, ,
\qquad f_0' = - \frac{\exp(E/T)}{(\exp(E/T) \pm 1)^2} \; \;
 \nonumber \\
d_t \delta & = & -  \left( \partial_t + {p_z\over E} \partial_z \right)
\left[ \mu + \mu_{bg} +p_z ( v + v_{bg} ) \right]
 - \frac{E}{T} \left( \partial_t +
\frac{p_z}{E} \partial_z \right) (\delta T +
\delta T_{bg}) \; \;
\end{eqnarray}
This equation is written in the fluid frame so that a particle's energy
is not conserved in the presence of a moving wall.  The term which
perturbs the population densities away from equilibrium  is
\begin{equation}
\partial_t E f_0'= {\partial_t m^2 \over 2E} f_0'
\label{eq:force term dot E}
\end{equation}
Note that this term is proportional to $m^2$.
This is the reason it is sensible to expand to linear order in perturbations.
First note that $m << T$.  For thermal
particles with $E \simeq T$, $-f_0' \simeq f_0$ and the source
for $\delta f/f$ is $O(m^2/ET)$
which is small.  For infrared fermions, $-f_0' \simeq f_0/2$ and
the source is $O(m^2/ET)$ which is also small.  For infrared bosons the
situation is a little less rosy; at small energy $-f_0' \simeq f_0 T/E$ and
the perturbation is $O(m^2/E^2)$.  Again, we will assume that these particles
are scattered efficiently and postpone a more careful treatment.

The Boltzmann equations have now become
\begin{eqnarray}
(-f_0') \left( \frac{p_z}{E} [ \partial_z ( \mu + \mu_{bg})
 +\frac{E}{T} \partial_z ( \delta T + \delta T_{bg} )
 + p_z \partial_z ( v + v_{bg}) ]
+   \partial_t (\mu + \mu_{bg} )
\right.  \nonumber \\
  \left. + \frac{E}{T} \partial_t ( \delta T + \delta T_{bg} )
 + p_z \partial_t ( v + v_{bg} )  \right)
 + C(\mu , \delta T ,
v) = (-f_0') \frac{\partial_t(m^2)}{2E}
\end{eqnarray}
The collision term depends on the deviations of all particle species from
the common background temperature and velocity.  We discuss it in some
detail in Appendix A.

The three parameters are determined uniquely by taking three integrals of
the Boltzmann equation; since our perturbations
are Lagrange multipliers for
particle number,
energy, and momentum, the appropriate choice is
$\int d^3 p/(2\pi)^3$, $\int E \: d^3 p/(2\pi)^3$, and $\int p_z \:
d^3 p/(2\pi)^3$.
The resulting equations are
\begin{eqnarray}
c_2 \partial_t(\mu + \mu_{bg})
 + c_3 \partial_t (\delta T+\delta T_{bg}) +{c_3 T\over 3} \partial_z
( v + v_{bg})
+ \int \frac{d^3p}{(2\pi)^3T^2} C[f] & = & \frac{c_1}{2T} \partial_t m^2
\nonumber \\
c_3 \partial_t(\mu + \mu_{bg})
 + c_4 \partial_t (\delta T+ \delta T_{bg})
+{c_4 T\over 3} \partial_z( v + v_{bg})
+ \int \frac{Ed^3p}{(2\pi)^3T^3} C[f] & = & \frac{c_2}{2T} \partial_t m^2
\nonumber \\
\frac{c_3}{3} \partial_z(\mu + \mu_{bg}) + \frac{ c_4  }{3} \partial_t
(\delta T + \delta T_{bg}) +
{ c_4  T\over 3} \partial_t( v + v_{bg})
+ \int \frac{p_z d^3p}{(2\pi)^3T^3} C[f] & = & 0
\label{eq:fluidequation}
\end{eqnarray}
The constants $c_i$ are defined by
\[
c_i  T^{i+1} \equiv \int E^{i-2}(-f_0')\frac{d^3p}{(2\pi)^3}
\]
For fermions they are $c_{1f} = \ln 2/2\pi^2$,
$c_{if} = (1-2^{1-i})i!\zeta_i/2\pi^2,$ $i>1$.  For bosons they are
$c_{1b} = \ln(2/m)/2\pi^2$, $c_{ib} = i! \zeta_i/2\pi^2$.  Here and throughout
 $\zeta_i$ is the Riemann zeta function evaluated at $i$.  We have worked to
lowest order in $m/T$ here, but only $c_{1b}$ and $c_{2b}$ possess $O(m/T)$
corrections, and these are small.

For a stationary wall all quantities are functions of $x=z+v_w t$ and the
stationary fluid equations are obtained by a simple substitution
\begin{equation}
\partial_t \delta_i\rightarrow v_w\delta_i '\,,\qquad
\partial_z \delta_i\rightarrow \delta_i '
\label{eq:temporal to stationary}
\end{equation}

The collision integral $\int C[f]$ depends on $\mu$, $\delta T$, and $v$ of
all heavy species, which couples the fluid equations.  However, for the
top quarks the collisions which contribute to order $\alpha_s^2$ arise only
from collisions with light quarks and gluons, and for $W$ bosons the
contributions at order $\alpha_s \alpha_w$ and $\alpha_w^2$ are primarily
with light quarks (the top quark making up only 1/6 of the population of
quarks).  Hence it is reasonable to ignore direct collisions between top
quarks and $W$ bosons.  If we included them then the problem would become
more complex, as these interactions mix top and bottom quarks,
and distinguish between top helicities.  To treat them properly we would need
to introduce separate fluid equations for left handed bottom quarks and for
each top quark handedness.  Because the effect is $\sim 10 \%$, we will
neglect it and treat all $W$ boson collisions with quarks as being with
light quarks.

The Higgs boson interacts predominantly with the top quark via its Yukawa
coupling, but since this interaction is quite efficient ($O(\alpha_t \alpha_s
\ln1/\alpha_s)$) and the Higgs has only one degree of freedom (and is lighter
than the $W$ boson if the phase transition is to be strong enough
for baryon number to be conserved after its completion) we will ignore Higgs
particles altogether.  Hence the collision terms in the fluid equations may
be treated as arising entirely through interactions with the background of
light particles.

We have computed the collision integrals appearing in
(\ref{eq:fluidequation}),
including all diagrams which contribute to order $\alpha_{s}^{2}
\ln 1/ \alpha_s$ for top quarks and to order  $\alpha_w^2
\ln 1/ \alpha_w$  for $W$ bosons, in Appendix A.
All diagrams are evaluated in the
{\it leading-log\/}  approximation. This means that only $t$-channel processes
(which are logarithmically divergent in the limit of the zero exchange particle
mass) are calculated.
The result for top quarks is
\begin{eqnarray}
\int \frac{d^3p}{(2\pi)^3T^2}C[f] = \mu \Gamma_{\mu1f}+ \delta T \Gamma_{T1f}
\qquad \Gamma_{\mu1f} =.00899 T \qquad \Gamma_{T1f} = .01752 T
\nonumber \\
\int \frac{d^3p}{(2\pi)^3T^3} E C[f] = \mu \Gamma_{\mu2f} +
\delta T \Gamma_{T2f} \qquad \Gamma_{\mu2f} =  .01752 T \qquad
\Gamma_{T2f} = .07172 T
\nonumber \\
\int \frac{d^3p}{(2\pi)^3T^3} \, p_z \, C[f] = vT \Gamma_{vf} \qquad
\Gamma_{vf} = .03765 T
\end{eqnarray}
We use the subscript $f$ (fermion) for top quarks and $b$ (boson) for $W$
bosons, but we suppress them when no confusion will occur.

For $W$ bosons we get
\begin{eqnarray}
\int \frac{d^3p}{(2\pi)^3T^2}C[f] = \mu \Gamma_{\mu1b}+ \delta T \Gamma_{T1b}
\qquad \Gamma_{\mu1b} = .00521 T \qquad \Gamma_{T1b} = .01012 T
\nonumber \\
\int \frac{d^3p}{(2\pi)^3T^3} E C[f] = \mu \Gamma_{\mu2b} +
\delta T \Gamma_{T2b} \qquad \Gamma_{\mu2b} = .01012 T \qquad
\Gamma_{T2b} = .03686 T
\nonumber \\
\int \frac{d^3p}{(2\pi)^3 T^3}\, p_z \, C[f] = vT \Gamma_{vb}
\qquad \Gamma_{vb} = .01614 T
\end{eqnarray}

The fluid equations then become
\begin{eqnarray}
v_w c_2 (\mu' + \mu_{bg}') + v_w c_3 (\delta T'+\delta T_{bg}')
 + \frac{c_3T}{3} (v'+ v_{bg}')
 + \mu \Gamma_{\mu1} + \delta T \Gamma_{T1}  & = &
\frac{v_w c_1}{2T} (m^2)'
\nonumber \\
v_w c_3 (\mu' + \mu_{bg}') + v_w c_4 ( \delta T' + \delta T_{bg}')
 + \frac{c_4 T}{3} (v'+ v_{bg}')
 + \mu \Gamma_{\mu2} + \delta T \Gamma_{T2} & = &
\frac{v_w c_2}{2T} (m^2)'
\nonumber \\
 \frac{c_3}{3} (\mu' + \mu_{bg}')
 + \frac{c_4}{3} ( \delta T' + \delta T_{bg}')
+ \frac{v_w  c_4  T}{3} ( v' + v_{bg}')
+ v T \Gamma_v & = & 0
\label{fluideq}
\end{eqnarray}
where for top quarks one should use $c_f$ and $\Gamma_f$ and for $W$ bosons
one should use $c_b$ and $\Gamma_b$.

We caution the reader not to interpret the $\Gamma$ simply as rates for
processes, without including the coefficients which appear in the derivative
terms of the fluid equations.  For instance, the rate at which a chemical
potential for top quarks decays is not
$\Gamma_{\mu 1 f} \sim T/110$, but roughly
$\sim \Gamma_{\mu 1f}/ c_{2f} \sim T/9$, corresponding to a typical lifetime
for a top quark before it annihilates with an  anti-top
of $18/T$ (the factor
of two is because each annihilation destroys two top type particles).
Similarly, the time it takes for successive small angle collisions to
randomize a particle's velocity is roughly $c_4 / 3 \Gamma_v \simeq 10/T$
for top quarks.

Collisions between massive species and light species appear in the fluid
equations of the light species with opposite sign.  Since the light species
are treated as being at a common temperature and velocity, no $\delta
T_{bg} \Gamma_{T}$ or $v_{bg} \Gamma_v$ appears (see Appendix A).  (If we
treated the background species separately, terms
order $\alpha_s^2 \ln 1/\alpha_s$ would damp the difference between background
species temperatures, velocities, and chemical potentials, efficiently forcing
them to equal.)  The background chemical potential is damped, but only by
inelastic processes which enter at $O(\alpha_s^3 \ln 1/\alpha_s)$.
The background fluid equations are
\begin{eqnarray}
\sum c_2 v_w \mu_{bg}' + \sum c_3 (v_w \delta T_{bg}' \! + \!
 \frac{T v_{bg}'}{3})
+ N_{bg} \Gamma_{\mu bg} \mu_{bg} = & & \nonumber \\
 N_t (\mu_f  \Gamma_{\mu1f} + \delta T_f \Gamma_{T1f} )
& + & N_W ( \mu_b  \Gamma_{\mu1b} + \delta T_b  \Gamma_{T1b} )
\nonumber \\
\sum c_3 v_w \mu_{bg}' +
\sum c_4 (v_w \delta T_{bg}' + \frac{T v_{bg}'}{3}) = & &  \nonumber \\
 N_t ( \mu_f   \Gamma_{\mu2f} + \delta T_f \Gamma_{T2f} )
& + & N_W ( \mu_b \Gamma_{\mu2b} + \delta T_b  \Gamma_{T2b} )
\nonumber \\
\frac{\sum c_3}{3} \mu_{bg}' +
\frac{\sum c_4}{3} ( \delta T_{bg}' + v_w T v_{bg}')
& = & N_t v_f T \Gamma_{vf} + N_W v_b T \Gamma_{vb} \quad
\end{eqnarray}
which simply state that the particle number, energy, and momentum
lost to the massive species
via collisions are taken up by the light species.  Here $\sum c_4 = 78 c_{4f}
+ 19 c_{4b} = (87.25)4\pi^2/30$, which is the heat capacity of the light
degrees of freedom, and $N_t = 12$ and $N_W = 9$ are the number of degrees
of freedom of top quarks and $W$ bosons.  The collision rate $\Gamma_{\mu bg}$
is the average over all background species of the particle number destruction
rate.

Let us estimate the importance of $\mu_{bg}$; consider a very thick wall so
that all derivatives may be dropped, and for simplicity ignore $W$ bosons and
$\delta T$ and $v$.  Then there are two fluid equations to consider,
\[
N_{bg} \Gamma_{\mu bg} \mu_{bg} = N_t \mu_t \Gamma_{\mu1} \qquad
\Gamma_{\mu1} \mu_t  = v_w\frac{c_1}{2T} (m^2)'
\]
The friction on the wall depends on the chemical potential of the top
quark, which is $\mu_t + \mu_{bg}$.  In this
approximation, it is
\[
v_w\frac{c_1}{2T} (m^2)'
 \left( \frac{1}{\Gamma_{\mu1f}} + \frac{N_t}{ N_{bg} \Gamma_{\mu bg}} \right)
\]
We see that $\mu_{bg}$ is important when
$N_t \Gamma_{\mu 1} > N_{bg}\Gamma_{\mu bg}$, or when the total rate at which
background particles convert into top quarks exceeds the total rate at which
they annihilate via inelastic processes.  If these inelastic processes were
very inefficient, the friction on the wall could be significantly
enhanced.  Unfortunately it is quite hard to compute $\Gamma_{\mu bg}$, as
there are many diagrams, all with 5 particles.  The diagrams involving
several gluons interfere and are not separately gauge invariant.  While
the diagrams are at a high perturbative order, they gain infrared log
enhancements when there is a $t$ channel exchange and one incoming gluon is
soft.  The diagrams involving gluons have very large
color factors because the gluons are in the adjoint representation, and Bose
statistics also make these diagrams large.  We have made a crude estimate of
these rates and find that
$N_{bg} \Gamma_{bg}$ is significantly larger than $N_t \Gamma_{\mu1f}$,
so we will neglect the background species chemical potential in what
follows.  (This conclusion sounds like an invalidation of the perturbative
expansion in $\alpha_s$, but this is not so: $N_{bg}$ is several
times larger
than $N_t$, and this makes up for the diagrams being at a higher perturbative
order.)

In this approximation the background fluid equations become rather simple
\begin{eqnarray}
\sum c_4 (v_w \delta T_{bg}' + \frac{T v_{bg}'}{3}) & = &
 N_t (\mu_f  \Gamma_{\mu2f} + \delta T_f \Gamma_{T2f} )
+ N_W ( \mu_b  \Gamma_{\mu2b} + \delta T_b  \Gamma_{T2b} )
\nonumber \\
\frac{\sum c_4}{3} ( \delta T_{bg}' + v_w T v_{bg}')
& = & N_t v_f T \Gamma_{vf} + N_W v_b T \Gamma_{vb} \,,\qquad
\mu_{bg}=0
\label{background fluideq}
\end{eqnarray}
Integrating these expressions, together with Eq. (\ref{fluideq}), reproduces
Eq. (\ref{boundary condition}).
The derivative term on the lefthand side
is exactly what appears in the fluid
equations for the perturbed species, which allows us to eliminate the
background perturbations in the fluid equations by direct substitution.
The final form of the fluid equations  for the top quarks is
\begin{eqnarray}
v_w c_{2f} \mu_f' + v_w c_{3f} \partial T_f' + \frac{c_{3f} T}{3}v_f '
+ \mu_f \Gamma_{\mu1ff} + \delta T_f \Gamma_{T1ff} + \mu_b \Gamma_{\mu1fb}
+ \delta T_b \Gamma_{T1fb}  =  F_{1f}
\nonumber \\
v_w c_{3f} \mu_f' + v_w c_{4f} \partial T_f' + \frac{c_{4f} T}{3}v_f '
+ \mu_f \Gamma_{\mu2ff} + \delta T_f \Gamma_{T2ff} + \mu_b \Gamma_{\mu2fb}
+ \delta T_b \Gamma_{T2fb}  =  F_{2f}
\nonumber \\
\label{fluid wo background}
\frac{c_{3f}}{3} \mu_f' + \frac{c_{4f}}{3} \delta T_f'
+ \frac{v_w  c_{4f}  T}{3}v_f ' + v_f T \Gamma_{vff} + v_b T \Gamma_{vfb}
  =  0 \; \; \;
\end{eqnarray}
where
\begin{eqnarray}
F_{1f} = \frac{v_w c_{1f}}{2}(m_t^2)' & \quad &
 F_{2f} = \frac{v_w c_{2f}}{2}(m_t^2)'
\nonumber \\
\Gamma_{\mu1ff} = \Gamma_{\mu1f} + \frac{N_t c_{3f}}{\sum c_4} \Gamma_{\mu2f}
& \quad &
\Gamma_{T1ff}= \Gamma_{T1f} + \frac{N_t c_{3f}}{\sum c_4} \Gamma_{T2f}
\nonumber \\
\Gamma_{\mu2ff} = (1 + \frac{N_t c_{4f}}{\sum c_4}) \Gamma_{\mu2f} & \quad &
\Gamma_{T2ff} = (1 + \frac{N_t c_{4f}}{\sum c_4}) \Gamma_{T2f}
\nonumber \\
\Gamma_{\mu1fb} =  \frac{N_W c_{3f}}{\sum c_4}  \Gamma_{\mu2b} & \quad &
\Gamma_{T1fb} =  \frac{N_W c_{3f}}{\sum c_4}  \Gamma_{T2b} \nonumber \\
\Gamma_{\mu2fb} =  \frac{N_W c_{4f}}{\sum c_4}  \Gamma_{\mu2b} & \quad &
\Gamma_{T2fb} =  \frac{N_W c_{4f}}{\sum c_4}  \Gamma_{T2b}  \nonumber \\
\Gamma_{vff} = (1 + \frac{N_t c_{4f}}{\sum c_4}) \Gamma_{vf} & \quad &
\Gamma_{vfb} = \frac{N_w c_{4f}}{\sum c_4} \Gamma_{vb}  \nonumber
\end{eqnarray}

The $W$ fluid equations look the same with the replacements
$b \leftrightarrow f$, $N_W \leftrightarrow N_t$.  The collision terms have
become (weakly) coupled between the heavy species, now indirectly through their
influence on the background.

We also extract the behavior of the background temperature for future use:
\begin{equation}
\delta T_{bg}' = \frac{ N_t ( \mu_f \Gamma_{\mu2f} +  \delta T_f \Gamma_{T2f})
+ N_W ( \mu_b \Gamma_{\mu2b} + \delta T_b \Gamma_{T2b} )}
{\sum c_4 (1/3 -v_w^3)}
\label{BGtempeq}
\end{equation}
We see
that the resistance to the wall's movement from the heating of the plasma
becomes important as the wall approaches the speed of sound.  The divergent
behavior at the speed of sound signifies the breakdown of our linearization
of perturbations.

\section{Delta function response}

In order to gain some intuition for the fluid equations
we  study the
response to a delta function source.  Consider the fluid equations for
the top quark, ignoring for the moment the change in the background
so there is no coupling to the $W$ boson.  We can write the fluid equations
in a matrix form,
\begin{equation}
A \delta ' +\Gamma \delta = F
\end{equation}
where
\begin{eqnarray}
\delta= \left ( \begin{array}{l}
 \mu  \\ \delta T \\ v \end{array} \right ) \,, & \qquad
     F=\left (
\begin{array}{l} F_1 \\ F_2     \\ 0 \end{array} \right ) \nonumber \\
V=\left ( \begin{array}{ccc} c_2 v_w & c_3 v_w & {c_3 \over 3} \\
                  c_3 v_w & c_4 v_w & {c_4 \over 3} \\
                  \frac{c_3}{3} & \frac{c_4}{3} & \frac{c_4 v_w}{3}
  \end{array}\right )
\,, & \qquad
\Gamma=\left ( \begin{array}{ccc} \Gamma_{\mu 1} & \Gamma_{T 1} &  0 \\
                       \Gamma_{\mu 2} & \Gamma_{T 2} &  0 \\
                          0   &    0  & \Gamma_v \end{array} \right )
\nonumber
\end{eqnarray}
Since the fluid equation is linear it  suffices  to study the solution
when the source is a delta function times some column vector $F$.  To solve
this
we need the homogeneous solution,
\[
\delta = \sum_i \alpha_i \chi_i \exp(-\lambda_i x) \, \qquad x = z+v_w t
\]
where $\alpha_i$ are coefficients, $\lambda_i$ are the solutions to
\begin{equation}
{\rm Det} [-A\lambda +\Gamma ]=0
\label{root equation det}
\end{equation}
and $\chi_i$ is the vector annihilated by $-\lambda_i A + \Gamma$.

To solve the problem with a delta function source, we write a general solution
to the homogeneous equations on each side of the source and apply boundary
and matching conditions.  Because $\delta$ should go to 0 at large distances,
only the positive values of $\lambda$ can have nonzero coefficients in
front of the source and only the negative values can have nonzero coefficients
behind.  We write
\begin{equation}
\delta = \left\{ \begin{array}{cc}
	\sum_{\lambda_i > 0} \alpha_i \chi_i \exp{-\lambda x} & x>0 \\
	\sum_{\lambda_i < 0} \alpha_i \chi_i \exp{-\lambda x} & x<0
	\end{array} \right.
\end{equation}
and determine $\alpha_i$ from $A \delta(0^{+}) - A \delta(0^{-} ) = F$,
the matching condition across the delta function.  This gives
\begin{equation}
\sum_i {\rm sign}(\lambda_i) \alpha_i A \chi_i = F
\end{equation}
which is solved by expanding $A^{-1}F$ in the eigenvalues $\chi$.

We see that the solution consists of several tails, some in front of the
source and some behind it.
These tails model the transport of the perturbation around the source due
to particle flow.

It is interesting to know how far
the particles spread and how asymmetric the spreading is, so we will
very briefly investigate the roots to Eq. (\ref{root equation det}).
This equation is a polynomial in $\lambda$.  The coefficients are very messy,
but much of what we want to know is in the coefficient for $\lambda^3$,
\[ \frac{c_4 v_w}{3} (\frac{1}{3} - v_w^2)(c_2 c_4 - c_3^2) \]
and the coefficient for $\lambda^0$,
\[ \Gamma_v (\Gamma_{\mu1} \Gamma_{T2} - \Gamma_{\mu2} \Gamma_{T1}) \]
Their ratio gives the product of the $\lambda$'s, and its sign tells us
how many of the tails precede the source and how many follow it, as a
function of $v_w$.  We immediately see that, for a subsonic wall, one
tail precedes the source and two follow it; but the sign of the coefficient
for $\lambda^3$ changes at $v_w = v_s = 1/\sqrt{3}$, and all three roots
then trail the source.  No diffusion occurs in front of a supersonic wall,
at least within the fluid approximation.  It is thus quite important to the
study of baryogenesis to know whether the bubble wall is subsonic or
supersonic.

We will also comment that, in the special case that $v_w =0$ and the
particle decay rate is very much slower than the scattering rate (so
$\Gamma_{T2}, \Gamma_v >> \Gamma_{\mu1}, \Gamma_{T1}, \Gamma_{\mu2}$), the
root equation for $\lambda$ simplifies; it is approximately
\[ -\frac{c_3^2}{9} \Gamma_{T2} \lambda^2 + \Gamma_v \Gamma_{T2} \Gamma_{\mu1}
= 0 \]
with roots $\lambda \rightarrow \infty$ (a  non-propagating
disturbance) and
$\lambda^2 = 9 \Gamma_v \Gamma_{\mu 1}/c_3^2$, which are decay tails.  The
length of the tail is
$c_3/(3 \sqrt{\Gamma_v \Gamma_{\mu1}})$.
By finding the small $v_w$ limit of the coefficient for $\lambda^1$ and
comparing the tails to the result of the diffusion equation
we find that the diffusion length is $D = c_3^2/(9 c_2\Gamma_v)$ \cite{JPT},
which for our value of $\Gamma_v$ is $D=2.7/T$ for top quarks and $D=5.5/T$
for $W$ bosons.

{}From the collision rates presented in the last section, we find the length
of the tails at $v_w = 0$ is $7.4/T$ for top quarks and $18/T$ for
$W$ bosons.  Top quarks do not spread very far, but $W$ bosons do.  However,
both lengths are larger than the diffusion lengths of the species, so the
journey is a random walk and not a free flight.

\section{Wall Velocity with a Wall Shape {\it Ansatz}}

\label{ansatzz}
In the fluid approximation the equation of motion of the Higgs field,
Eq. (\ref{eqmo}), becomes
\begin{equation}
-(1-v_w^2)\phi'' + V_T'(\phi , T) + \frac{N_tT}{2}\frac{dm_t^2}{d\phi}
(c_{1f} \mu_f + c_{2f} \delta T_f)+
 \frac{N_W T}{2} \frac{dm_W^2}{d\phi} (c_{1b}\mu_b + c_{2b} \delta T_b) = 0
\label{fluideqmo}
\end{equation}

This, together with the fluid equations and the equation for the background
temperature, constitute a well posed set of equations for
the shape of the wall.  They are velocity dependent; we
expect them to have a solution
at a discrete set of velocities (hopefully one).  However, they are
nonlinear, so their solution must be numerical.  For starters it would be
useful to see how much progress we can make analytically.  We have done
so in an earlier paper \cite{paper1}; here we will improve that analysis
by including the background temperature.

Let us restrict the shape of the wall to an {\it Ansatz},
\begin{equation}
\label{Ansatz}
\phi(z,t) = \frac{\phi_0}{2} \left( 1 + {\tanh}\frac{z+v_w t}{L} \right)
\end{equation}
where $v_w$ and $L$ are parameters to be optimized.  If Eq. (\ref{fluideqmo})
were derived from a free energy $F$ we would know how
to proceed: we would solve $\partial F/\partial L = \int {\rm Eq.}
(\ref{fluideqmo}) \partial\phi/ \partial L=0$ and $\partial F/\partial v_w =
\int {\rm Eq.}
(\ref{fluideqmo}) \partial\phi/ \partial v_w=0$ simultaneously.
Since our equation of motion is dissipative, there is no free energy which
generates it; however, these constraints still
have the right physical content. Noting that
$\partial \phi/ \partial L = -(z+v_wt) \phi' /L$ and $\partial \phi/
\partial v_w = t \phi'$, we guess that a good pair of constraints should be
\begin{equation}
\int ({\rm Eq.}\ref{fluideqmo}) \phi' dz = 0
\qquad
\int ({\rm Eq.}\ref{fluideqmo}) \frac{z}{L} \phi' dz  = 0
\label{constraints}
\end{equation}
Indeed, these have sensible physical interpretations.  From Eq.
(\ref{Newtonforphi}), we see that the
first of these constraints is that the total pressure on the wall should be
zero; if the
total pressure were nonzero then the wall would accelerate, changing $v_w$.
The second equation is an
asymmetry in the total pressure between the front and back of the wall.  If
it were not zero there would be a net compressive or stretching force on
the wall, changing $L$.

Next we must deal with the variation of temperature across the wall.  Write
the temperature in the broken phase as $T$ (which is $T_a$ in the language
of section \ref{hydrosec})
 and the temperature at a position
$z$ as $T(z) = T + \delta T_{bg}(z)$.  We will solve for $\delta T_{bg}$
using Eq. (\ref{BGtempeq}) and the
boundary condition $\delta T_{bg}(z \rightarrow \infty) = 0$ and correct
(\ref{fluideqmo}) to linear order in $\delta T_{bg}$.  The correction is
$\delta T_{bg} dV_T'/dT \simeq 4D\phi T \delta T_{bg}$.

The integrals (\ref{constraints}) for the constant temperature, equilibrium
part of (\ref{fluideqmo}) can be performed; they give
\begin{eqnarray}
 \int  ( \Box \phi + V_{T}'(\phi)) \phi' & = & V_T (\phi_0) - V_T
(0)
\equiv - \Delta V_T \label{eq:constraint one}\\
 \int  [ \Box \phi + V_{T}'(\phi) ] \frac{z}{L} \phi' & = &
\frac{(1-v^2_w) \phi^2_0}{6L^2} - \frac{1}{2}[ \Delta V_T + \Xi
]
 \label{eq:constraint two}\\
 \Xi  \equiv   C\phi_0^2
(\zeta_2 - 1) T^2 & + & \frac{E\phi_0^3 T}{2}  - \frac{5\lambda_T
\phi_0^4}{24}
   + \frac{A_f g_w^6 T^4}{12}\left (2.79 + {1\over 2} \ln
{\phi_0\over
T}\right ) \label{eq:constraint three}
\end{eqnarray}
Note that the $\Box \phi$ term acts to stretch the wall (increase
$L$) while $V_T$ acts to accelerate and
compress the wall.
The coefficient 2.79 in the last term is the only place where
our choice for the function Pit enters our computation.

Next we must evaluate the integrals
\[
4DT \int \delta T_{bg} \phi \phi' dz \, , \quad
4DT \int \delta T_{bg} \frac{z}{L} \phi \phi' dz
\]
for the background temperature,
\[
\int N_t (c_{1f} \mu_f + c_{2f} \delta T_f)
   \frac{y^2}{2} \phi \phi'(z) dz \, , \quad
\int N_t (c_{1f} \mu_f + c_{2f} \delta T_f)
   \frac{y^2}{2}  \frac{z}{L} \phi \phi'(z) dz
\]
for quarks, and similarly for W bosons; to do this we Fourier transform
the integrals to
\[
\int N_t (c_{1f} \tilde{\mu}_f (k) + c_{2f} \widetilde{\delta T}_f (k))
\frac{y^2}{2}  \widetilde{\phi \phi'}(-k) \frac{dk}{2\pi}
\]
et cetera.  One complication is that $c_{1b}$ is weakly $z$ dependent;
we approximate
it at its value where $\phi \phi'$ is maximum.  Evaluating
$\widetilde{\phi\phi'}$ by a contour integration, we find
\begin{equation}
\widetilde{\phi\phi'}(k) = \frac{\phi_0^2}{2}
(1 - ikL/2) \frac{kL\pi}{2} {\rm csch}\frac{kL\pi}{2} \qquad
\frac{\widetilde{z\phi\phi'}(k)}{L} = i \frac{d\widetilde{\phi\phi'}}{d(kL)}
\end{equation}
We determine $\tilde{\mu}$ and $\widetilde{\delta T}$ from the fluid equations;
they can be written in a matrix form by writing
$\delta = [ \mu_f,\, \delta T_f,\,v_f,\, \mu_b ,\, \delta T_b ,\, v_b ]^T$;
the fluid equations become
\begin{equation}
A_{ij} \delta_j' + \Gamma_{ij} \delta_j = F_i \phi \phi' \qquad
\delta T_{bg} ' = R_i \delta_i
\label{matrixfluideq}
\end{equation}
The coefficient  matrices  $A$ and $\Gamma$ and the source vector $F$ can
be read off from Eq. (\ref{fluid wo background}),
and the form of $R$ can be read off from Eq.(\ref{BGtempeq}).

$F_i$ has one term $\sim c_{1b}$ which is weakly
position dependent; again we approximate it as its value where $\phi\phi'$
is maximum.  The fluid equations are then easy to Fourier transform,
becoming
\begin{equation}
ik\tilde{\delta_i} + (A^{-1})_{ij} \Gamma_{jk} \tilde{\delta}_k
 = (A^{-1})_{ij} F_j \widetilde{\phi\phi'}
\qquad ik \widetilde{\delta T}_{bg} = R_i \tilde{\delta}_i
\label{eq1fordelta}
\end{equation}
Now denote the eigenvalues of $(A^{-1}\Gamma)$ as $\lambda_i$, and write
the matrix whose columns are the eigenvectors of $(A^{-1}\Gamma)$ as
$\chi$, so that
\[
(A^{-1}\Gamma)_{ij}\chi_{jk} = \chi_{ik} \lambda_k
\]
 (with
no sum on $k$).  Also define $\alpha_i = (\chi^{-1})_{ij}\delta_j$ and
$S_i = (\chi^{-1})_{ij}(A^{-1}F)_j$, which are the deviation
from equilibrium and the source in the basis of eigenvectors.  Multiplying
the fluid equations on the left by $\chi^{-1}$, we find
\begin{equation}
\tilde{\alpha}_i = \frac{S_i}{\lambda_i + ik} \widetilde{\phi \phi'}(k)
\qquad ik \widetilde{\delta T}_{bg} = R_i \chi_{ij} \frac{S_j}{\lambda_j + ik}
\widetilde{\phi \phi'}(k)
\end{equation}

The equation for $\widetilde{\delta T}_{bg}$ only
determines its value up to a constant
of integration, which in Fourier space is a delta function; the coefficient
of the delta function is determined by the boundary condition for $\delta
T_{bg}$.  We find
\begin{equation}
\widetilde{\delta T}_{bg} = R_i \chi_{ij} \left( \frac{S_j}{ik(\lambda_j
+ik)} - \frac{S_j \pi}{\lambda_j}\delta(k) \right)
\widetilde{\phi \phi'}(k)
\end{equation}
where $\delta(k)$ is the Dirac delta function.

The contributions to Eq. (\ref{constraints}) in this notation are
\begin{equation}
\int \frac{dk}{2\pi} \left[ f_i \chi_{ij} \frac{S_j}{\lambda_j + ik}
+ 4DT R_i\chi_{ij} \left( \frac{S_j}{ik(\lambda_j + ik)}- \frac{S_j \pi}
{\lambda_j} \delta (k) \right) \right] \widetilde{\phi \phi'}(k)
\widetilde{\phi\phi'}(-k)
\label{horrorhorror1}
\end{equation}
and
\begin{equation}
\int \frac{dk}{2\pi L} \left[ f_i \chi_{ij} \frac{S_j}{\lambda_j + ik}
+ 4DT R_i\chi_{ij} \left( \frac{S_j}{ik(\lambda_j + ik)}- \frac{S_j \pi}
{\lambda_j} \delta (k) \right) \right] \widetilde{\phi \phi'}(k)
\widetilde{z\phi\phi'}(-k)
\label{horrorhorror2}
\end{equation}

The vector $f$ above, which gives the force on the field from $\delta$,
can be read off from Eq. (\ref{fluideqmo}).

How does transport enter these integrals?  The $\lambda_i$ are the inverse
lengths of exponential tails.  The  no transport  limit
is the limit in which the
$\lambda_i$ are large.  In this limit the $ik$ in the denominator is
irrelevant.  Since $\widetilde{\phi \phi' }(k) \widetilde{\phi \phi '}(-k)$
is real, the $ik$ reduces
the value of Eq. (\ref{horrorhorror1}); transport, by
spreading out the perturbation, has reduced the friction on the wall.

Eq. (\ref{horrorhorror1})- (\ref{horrorhorror2})
may be solved using the following integrals:
\begin{eqnarray}
\int \frac{dk}{2\pi} \frac{1}{\lambda + ik} \widetilde{\phi\phi'}(k)
 \widetilde{\phi\phi'}(-k) & = & \frac{\phi_0^4}{16} \left( (\lambda L -\frac{
(\lambda L)^3}{4} ) I_1({\lambda L\pi\over 2}) + \frac{\lambda L}{3}\right)
\label{frictioneq}\\
\int \frac{dk}{2\pi L} \frac{1}{\lambda + ik} \widetilde{\phi\phi'}(k)
 \widetilde{z \phi\phi'}(-k) & = & \frac{\phi_0^4}{16}
 \left( ( \frac{(\lambda L)^2}{2} + \frac{\lambda L}{2} -1 )
I_1(\frac{\lambda L \pi}{2}) \right. \nonumber \\
 & & \left.   + (1 - \frac{(\lambda L)^2}{4} )
I_2({\lambda L \pi \over 2}) - \frac{1}{6} \right)
\label{stretcheq} \\
\int \frac{dk}{2\pi} \frac{1}{ik(\lambda + ik)} \widetilde{\phi\phi'}(k)
\widetilde{\phi\phi'}(-k) & = & \frac{-1}{\lambda}
({\rm Eq.} \ref{frictioneq})
\\
\int \frac{dk}{2\pi L} \frac{1}{ik(\lambda + ik)} \widetilde{\phi\phi'}(k)
 \widetilde{z \phi\phi'}(-k) & = & \frac{-1}{\lambda}
( {\rm Eq.} \ref{stretcheq})
- \frac{5\phi_0^4}{96} \; \; \;
\end{eqnarray}
Where we have defined the integrals
\begin{equation}
I_1(a) = \int_{-\infty}^{+\infty} \frac{x^2 {\rm csch}^2x}{x^2 + a^2} dx
 \qquad
I_2(a) = \int_{-\infty}^{+\infty} \frac{x^3 {\rm csch}^2x {\rm coth}x}
{x^2 + a^2} dx
\end{equation}

Evaluating these by contours gives
\begin{eqnarray}
I_1(a)  & = & \frac{\pi a}
{\sin^2 a} - 2 - \sum^{\infty}_{n=1} \frac{n (2\pi a)^2}
{( (n\pi)^2 - a^2)^2} \nonumber \\
I_2(a) & = & \frac{\pi a^2 \cos a}{\sin^3 a} - 1 + \sum^{\infty}_{
n = 1} 2n\pi^2 a^2 \frac{3 a^2 + (n\pi)^2}{((n\pi)^2 -a^2)^3}
\end{eqnarray}

Numerically, these expressions suffer from cancelling divergences near
$a=n\pi$.

For large values of $a$ the integrals possess useful asymptotic series
\begin{equation}
I_1(a) \rightarrow 4\sum_{n=1}^{\infty} \frac{(-)^{n+1} (2n)! \zeta(2n)}
{(2a)^{2n}} \qquad
I_2(a) \rightarrow 2\sum_{n=1}^{\infty}\frac{(-)^{n+1} (2n+1)! \zeta(2n)}
{(2a)^{2n}}
\end{equation}

At small values of $a$ it is useful to rearrange
the infinite series into Taylor expansions,
\begin{eqnarray}
I_1(a) & = & \frac{\pi a}{\sin^2 a} - 2 -4\sum_{n=1}^{\infty} n
 \left(\frac{a}{\pi} \right)^{2n} \zeta(2n+1) \nonumber \\
I_2(a) & = & \frac{\pi a^2 \cos a}{\sin^3 a} -1 + \sum_{n=1}^{\infty}
n(n+1)\left(\frac{a}{\pi}\right)^{2n} \left( \frac{3a^3}{\pi^2}\zeta(3+2n)
+\zeta(1+2n)\right)
\end{eqnarray}
which have radius of convergence $\pi$.

This completes the evaluation of
Eq. (\ref{constraints}).  These two constraint
equations determine curves in the space of $v_w$, $L$ and their intersections
are self-consistent solutions for the wall shape and velocity within
our {\it Ans\"{a}tze} and approximations.  These curves are illustrated
in Figure \ref{vLmatch}.

\section{Solving the Equations of Motion}

It is also possible to solve (\ref{fluideqmo}) and the fluid equations
numerically for a general wall shape.  Although they constitute a nonlinear
system of equations, they are linear in $\delta$; we can
therefore solve for $\delta$ as (nonlocal) functions of $\phi$
and reduce the system to a single integro-differential relation for $\phi$.
We begin with Eq. (\ref{eq1fordelta}), but including the position dependence
of $F$, so we must Fourier transform $F\phi \phi'$ as a unit.
Now write
\begin{equation}
\tilde{S}_i \equiv (\chi^{-1})_{ij}(A^{-1})_{jk} \widetilde{F_k\phi \phi'}
\end{equation}

We quickly find
\begin{equation}
\tilde{\delta}_i = \chi_{ij} \frac{ \tilde{S}_j}{\lambda_j + ik}
\end{equation}
which can be inverse transformed into a convolution,
\begin{eqnarray}
\delta_i(z) = \chi_{ij} \int dy G_j(y-z) (\chi^{-1}A^{-1}F)_j(y)
\phi \phi'(y) \nonumber \\
G_j(x) \equiv {\rm sign}(\lambda_j) \theta(x \,{\rm sign}
(\lambda_j)) \exp(-\lambda_j x)
\end{eqnarray}
Here $G$ is the Green's function for fluid perturbations.  Now given a
spatial configuration for $\phi$, say on a lattice of points, we may integrate
numerically to find $\delta$ and integrate again to find $T_{bg}$.  This
gives us the full equation of motion (\ref{fluideqmo}).  If the equation
of motion had been derived from a free energy, then we know that changing
$\phi$ in the direction dictated by the equation of motion reduces the free
energy, and is guaranteed to approach a minimum if one exists.
We may hope that evolving $\phi$
in the direction dictated by the equation of motion will lead us towards a
solution of the equations of motion, but we must be a little careful because
there is only a solution at select values of $v_w$, and because the wall
has a zero mode.
The naive way to evolve $v_w$ towards its correct value is to move it
in the direction dictated by the total pressure on the wall, $\int
({\rm Eq.} \ref{fluideqmo}) \phi' dz$.
We then subtract a quantity proportional to
$\phi'$ from the equation of motion when we correct $\phi$,
as otherwise part of the correction we implement will just be a shift
in the wall position.  This approach should work below the speed of sound,
but above the speed of sound the pressure may decrease with velocity (as
the temperature elevation reduces).

We have implemented this procedure numerically, and find that the wall does
approach a solution of the equation of motion for subsonic starting velocity.
Since the lattice on which $\phi$ is defined need only be 1 dimensional (as
the wall is planar) it is easy to make the stepsize small enough that the
results are robust; our results agree to 0.2 \% when we change stepsize
from $1/T$ to $2/T$.  We find that the wall is somewhat deformed from a
tanh, appearing steeper on the side facing the symmetric phase and shallower
on the side facing the broken phase.  Part of this is due to the two loop
terms in the effective potential, but it also turns out to be true when we
use just the 1 loop potential, which produces a symmetric wall at equilibrium.
We can understand this deformation as follows; the source for excess particles
goes as $\phi \phi'$, which peaks above $\phi = \phi_0/2$.  The excess
 particles
are on average swept further up the wall towards large $\phi$ because of the
motion of the wall with respect to the plasma.  Hence most of the frictive
force on the wall occurs at large $\phi$, and stretches out the upper part
of the wall.


\section{Results and Conclusions}

The wall velocity computed from the tanh {\it Ansatz} and numerically are
tabulated for several effective potential parameters in
Table \ref{velocitytable}
\begin{table}
\begin{tabular}{|| c | c | c | c | c | c | c ||} \hline
$\lambda_T$ & $m_H(T=0)$ &
 $A_F$ & {\it Ansatz} $v_w$ & {\it Ansatz} $L$ & numerical $v_w$
& $\phi_0/T$  \\ \hline \hline
.0204& 0  & 0 & .351 & 29 & .365 & .987 \\ \hline
.023 & 34 & 0 & .356 & 28 & .374 & .907 \\ \hline
.03  & 50 & 0 & .365 & 26 & .392 & .757 \\ \hline
.04  & 70 & 0 & .377 & 25 & .412 & .635 \\ \hline
.05  & 81 & 0 & .390 & 24 & .428 & .562 \\ \hline
.06  & 91 & 0 & .401 & 23 & .441 & .516 \\ \hline
.03  & 50 & .1& .481, run & 18 & .496 & .998 \\ \hline
.04  & 70 & .1& .497, .97 & 15, 14 & .513 & .861 \\ \hline
.05  & 81 & .1& .508, .91 & 13, 11 & .524 & .777 \\ \hline
.03  & 50 & .2& .499, run & 15 & .510 & 1.12 \\ \hline
.04  & 70 & .2& .513, run & 12 & .524 & .971 \\ \hline
.05  & 81 & .2& .522, .98 & 11, 7.7 & .533 & .878 \\ \hline
.06  & 91 & .2& .528, .96 & 9.6, 6.2 & .539 & .814 \\ \hline
.04  & 70 & .3& .519, run & 11& .530 & 1.05 \\ \hline
.05  & 81 & .3& .527, run & 9.4& .537 & .952 \\ \hline
.06  & 91 & .3& .534, run & 8.5& .543 & .882 \\ \hline
.07  & 98 & .3& .538, run & 7.8& .546 & .831 \\ \hline
\end{tabular}
\caption{Wall velocity and thickness and Higgs vev after
transition at several effective potential parameters.  The second entries in
the {\it Ansatz} velocity and length columns for some entries are detonation
solutions; in some cases the {\it Ansatz} predicts runaway.}
\label{velocitytable}
\end{table}
 and the
wall shape they give is compared in Figure \ref{Blah}.
We see that the {\it Ansatz}
returns the velocity accurately, although at large $A_F$ it gives a wrong
wall shape.  The column $\phi_0/T$ is the Higgs vev after the phase transition
has completed, accounting for the heating of the plasma due to the release
of latent heat in the transition.

There is always a subsonic solution
within the approximations that we have made, because as one approaches the
speed of sound from below the temperature is elevated by more and more.
In all cases in the table with $A_F = 0$, the only solution is subsonic.
This is because the liberated latent heat raises the temperature of the
plasma, which inhibits the motion of the wall.  Neglecting the background
temperature, as we did in \cite{paper1}, produces a supersonic wall, so the
effect is quite important.  The result that the wall is subsonic is robust in
the sense that, if we have underestimated all collision rates by a factor
of two, we still find only a subsonic solution.  Also note that the velocity
is quite weakly dependent on the Higgs mass; in the range $0 < m_H < 90$Gev,
$.38< v_w < .46$.  Recall that this result ignores the Higgs
particle contribution to friction and may underestimate the contribution of
infrared $W$ bosons, which can only make the wall velocity lower.
However, top quarks were typically responsible for about $60 \%$ of the
friction and $65 \%$ of the liberated latent heat, so we do not anticipate
that infrared particles will change our results a great deal.

Including the background temperature has also reduced the dramatic stretching
of the wall found in \cite{paper1}.  This stretching arose because the
force on the wall from $\delta f$ is predominantly far back on the wall.
In the limit where decay rates are fast compared to the wall passage time,
$\delta f \sim \phi \phi'$, which peaks well above $\phi = \phi_0/2$.
The force on the wall from $\delta f$ depends on $\delta f \phi \phi'$,
which peaks even more strongly on the upper part of the wall.  Transport
compounds this effect as particles sweep up the wall because of its movement.
The result is that most of the force from $\delta f$ is far back on the
wall, stretching it out. However, $\delta T_{bg}$ is largest
(for subsonic walls) in front of the wall in the symmetric phase, so it
tends to exert more force on the front of the wall, which compresses it.
This partially compensates the stretching we found in \cite{paper1}.

In all cases in the table
in which $A_F \neq 0$, the {\it Ansatz} technique tells us that
there is a supersonic solution; this is always very relativistic and
sometimes accelerates without limit.
While this result is not robust (it may be incorrect
if friction from infrared particles is important),
it is possible that the wall does become supersonic in these cases.
However, the fluid approximation is not
valid in these conditions, because the front part of the wall
becomes quite thin. A more careful analysis shows that
the compression of the front of the wall greatly increases the
friction and prevents ultrarelativistic motion.
This is discussed in Appendix C.

In conclusion, if there is a gauge boson condensate in the symmetric
phase, and if neither this condensate nor infrared bosons impart substantial
friction, then the wall becomes supersonic.
On the other hand, if there is no gauge boson condensate in the symmetric
phase, the wall is definitely subsonic just due to friction from thermal
particles and the release of latent heat.

\section{Acknowledgements}

We thank the Sir Isaac Newton Institute for Mathematical Sciences, Cambridge,
England, for hospitality during the early part of this work.
We are very grateful to Misha Shaposhnikov, Arthur Heckler, Michael Joyce
and Neil Turok for enlightening conversations.  We would also like to
thank Marco Moriconi for reminding us how to do partial fractions.
GM acknowledges support from an
NSF graduate fellowship and TP support by PPARC.

\section{Appendix A:  Scattering Rates}

In this appendix we discuss the computation of collision integrals.
The collision integral appearing in the Boltzmann equation is, to second
order in $\alpha$,
\begin{equation}
C[f]  = \sum \frac{1}{2E_p}\int \frac{d^3k \; \; d^3p' \; \; d^3k'}
{(2\pi)^9 \; 2E_k \; 2E_{p'} \; 2E_{k'}}
|{\cal M}(s,t)|^2(2\pi)^4 \delta^4(p+k-p'-k') {\cal P} [f_i]
\label{collisionterm}
\end{equation}
\[
 {\cal P} [f_i] =
f_1f_2(1\pm f_3)(1\pm f_4) - f_3 f_4 (1\pm f_1)(1\pm f_2)
\]
where the sum is over all four leg diagrams, $p$ refers to the incoming
particle, $k$ is the particle it hits, and $p'$ and $k'$ are the outgoing
particles; the legs are labelled as $1,2,3,4$ respectively.
$\cal M$ is the scattering amplitude for the process. The $f$'s are
population factors; the positive expression represents a particle being
removed from the state with momentum $p$ by a collision and is weighted by
the population of particles in that state and of the state it collides with.
The negative term in ${\cal P}[f_i]$
accounts for particles scattering into the state.
The factors $1 \pm f$ for the outgoing particles arise from particle
statistics; the $\pm$ is a $-$ for fermions (corresponding
to Pauli blocking) and a $+$ for bosons (for stimulated emission).
Weldon has given an excellent derivation of this expression from the
discontinuity on the real time axis of the self-energy of the
propagator \cite{Weldon}.  Extending his technique to include gauge particles
in a general covariant gauge
proves to introduce difficulties.  The structure of the collision integral
arises from the poles of propagators in a self-energy diagram, and the
propagator $D_{\mu\nu}=(g_{\mu\nu} - \xi k_{\mu}k_{\nu}/k^2)/k^2$ has extra
on shell divergences for $\xi \neq 0$.  Such infrared problems have been
discussed by Braaten and Pisarski \cite{Pisdamprate}, who have shown that
it is sufficient to work in the gauge $\xi = 0$.

Let us begin by analyzing the population factors.  Without loss of generality
we may write $f_i = 1/( \exp{a_i} \pm 1)$, and noting that $(1 \pm f_i) =
f_i \exp{a_i}$ we find
\[
f_1 f_2 (1 \pm f_3)(1 \pm f_4) - (1 \pm f_1) (1 \pm f_2) f_3 f_4 =
(e^{a_3+a_4} - e^{a_1+a_2} ) f_1 f_2 f_3 f_4
\]
The equilibrium value of $a_i$ is
$\gamma (E_i - \vec{v} \cdot \vec{p}_i)/T$ with
a common $\vec{v}$ and $T$ for all species. In this case,
$C[f]=0$ by energy and momentum conservation.
This holds for higher order graphs as well.  In the Boltzmann approximation
the collision integral is local, so it also holds for a common
spatially varying temperature and velocity.

Now let
$a_i = \gamma (E_i  - \vec{p}_i \cdot \vec{v}_{bg})/T_{bg} - \delta_i$, with
$\delta_i$ small;  then
\begin{eqnarray}
\exp(a_1+a_2) & = &
\exp( \sum_i \gamma (E_i - \vec{p}_i \cdot \vec{v})/T_{bg}) \exp(
-\delta_1 - \delta_2)
\nonumber \\
& \simeq & \exp( \sum_i \gamma (E_i - \vec{p}_i \cdot \vec{v})/T_{bg})
(1 - \delta_1 - \delta_2) \nonumber
\end{eqnarray}
to linear order in $\delta$, and hence
\[
{\cal P}[f_i] \simeq
(\sum \delta) f_1 f_2 (1 \pm f_3) ( 1 \pm f_4)
\]
where the sum on delta is over incoming {\it minus} outgoing legs.

Now let us compute the $O(\alpha_s^2)$ collision rates for top quarks.
All $t$-channel diagrams which contribute at order $\alpha_s^2$ are shown in
the first column of Figure \ref{diagrams}.
There are also $s$-channel processes.  These diagrams interfere, but
luckily the cross-amplitudes are all $O(m^2/T^2)$ and can be neglected.
Top quarks can also decay directly, with a matrix element $O(g_w^2)$, but
this process is time dilated and
phase space suppressed, which reduces its importance by $O(m^2/ET)$. It
therefore contributes at order $\alpha_w\alpha_y$ and can be neglected.

When we compute the cross-sections of these diagrams using free particle
propagators, we find that the $t$-channel processes lead to logarithmic
infrared divergences.  At finite temperature these divergences are cut off
by the interaction of the exchange particle with the plasma, so the
$t$-channel diagrams contribute at order $\alpha_s^2 \ln 1/\alpha_s$.
The exact thermal propagators are quite complicated, but if we are willing
to make an error in the leading constant (but get the coefficient
of the log right)
we can make a simple approximation which renders the computation more
tractable.  For an exchanged (top) quark, the correct dispersion relation
is very similar to that of a massive particle with $m_q^2 = g_s^2 T^2/6$
\cite{WeldonFermion}.  This correction is of order the (space dependent)
contribution from the physical mass, $m_t^2 = y_t^2 \phi^2/2$, which we will
therefore neglect, making the collision integral $\phi$ independent.
For the bosons the dispersion relations are quite complex.  For space-like
momenta, such as occur for $t$-channel gluon or $W$ boson exchange,
the longitudinal components of the propagator are Debye screened, and the
transverse parts Landau damped, below the plasma mass, which is
$m_g^2= 2g_s^2T^2$ for gluons and $m_W^2 = (5/3)g_w^2 T^2$ for $W$ bosons
\cite{WeldonBoson}.
To approximate this effect we modify the denominator of the propagator,
making it $t-m^2$, where $m$ is the appropriate plasma mass of the
exchange particle. This
approximation is quite rough; it should yield the correct leading log
coefficient, but not the correct leading constant.  Hence we can only work
to leading log accuracy.
We therefore drop $s$ channel processes, which
do not produce logarithms, set $u = -s$ (which is legitimate in the leading log
approximation), and
keep only the $st/t^2$ dependence of annihilation and
absorption-reemission processes and the $s^2/t^2$ dependence of scattering
processes.

We will also systematically drop terms of order $m/T$, which allows us to
treat the outgoing lines as approximately massless.  In all cases the
collision rates are dominated by thermal particles, and the infrared log
divergence arises primarily when the exchange momentum is between $T$ and
$m$, so this approximation is justified.

\centerline{\bf Top quark annihilation rates}

Now we will compute a simple annihilation diagram, quarks go to gluons
with matrix element (in the leading log approximation)
$\simeq -(64/9)g_s^4st/(t-m_q^2)^2$. The collision integral,
integrated over $d^3 p/(2 \pi)^3 T^2$, is
\begin{eqnarray}
\frac{64g_s^4}{9}\int  \frac{d^3p}{(2\pi)^3 T^3 2E_p}\frac{d^3k}{(2\pi)^32E_k}
  [2\mu + (E_p+E_k) \delta T/T + (p_z + k_z) v ] \,
  \nonumber \\
\int \frac{d^3p'}{(2\pi)^32E_{p'}}\frac{d^3k'}{(2\pi)^32E_{k'}}
 \frac{-st}{(t-m_q^2)^2} (2\pi)^4 \delta^4(p+k-p'-k')  {\cal P}[f_i]
\end{eqnarray}
In the leading log approximation the energy transfer is small, so we may
approximate
\[
{\cal P} =
f_p f_k (1+f_{p'})(1+f_{k'}) \simeq
f_p f_k (1+f_p)(1+f_k)
\]
where the $1+f$ use the Bose population function and the $f$ use the Fermi
population function.
In this approximation, ${\cal P}$ is independent of $p'$ and $k'$ and the
integrals over $p'$ and $k'$ are Lorentz invariant and may
 be performed in the center of
mass frame.  The integral over $k'$ performs the three spatial delta functions,
so that $\vec{k'} = -\vec{p'}$.
The integral over $p'$, making a small mass approximation for $t$, is
\begin{equation}
\int \frac{p'^2 dp' d\Omega_{p'}}{(2\pi)^3 2 E_{p'} 2E_{k'}} 2\pi
\delta(2E_p - 2E_{p'})
\frac{(2p\cdot k)2p p' (1-\cos\theta ')}{(2p p' (1-\cos\theta ')+m_q^2)^2}
\simeq\frac{1}{8\pi}\ln\frac{2p\cdot k}{m_q^2}
\end{equation}
to leading log accuracy.
We now perform the remaining integrals in the plasma frame.
The argument of the log contains the only dependence on $p$ or $k$;
$2p \cdot k = 2|\vec{p}| |\vec k|(1-\cos \theta) + O(m^2_q)$, with
$\theta$ the plasma frame angle between $\vec{p}$ and $\vec{k}$.  The
remaining integrals are
\begin{eqnarray}
\frac{64g_s^4}{9 T^3}\frac{1}{8\pi}\int \frac{p^2 dp d\Omega_p}{(2\pi)^32E_p}
\frac{k^2dk d\varphi d \cos \theta}{(2\pi)^32E_k} \left[
2 \mu + (E_p + E_k)\delta T/T + (p_z + k_z)v \right ]
\nonumber \\
 \qquad
f_p f_k (1+f_p)(1+f_k) \ln\frac{2pk(1-\cos \theta)}{m_q^2}
\end{eqnarray}
The integral over $(p_z + k_z)v$ vanishes when we integrate over $d\Omega_p$,
although (only) it will contribute when we integrate over $p_z d^3p/T^3$.
The integral over $\theta$ gives approximately $2 \ln(4pk/m_q^2)$, again
dropping a constant and keeping only the leading log.  The remaining angular
integrals are all trivial.  The energy integrals are dominated by thermal
energies where we are justified in approximating $p = E_p$, and we get
\begin{equation}
\frac{g_s^4}{18\pi^5 T^3} \int pk [2\mu+(p+k)\delta T/T] \ln\frac{4pk}{m^2}
 f_p f_k (1+f_p)(1+f_k) dp dk
\end{equation}

Using the approximation
\begin{equation}
\int p^n \ln \frac{p}{T} f_p(1+f_p) dp \simeq
\ln(n+ 1/2) \int p^n f_p (1+f_p) dp
\label{leadinglog}
\end{equation}
which is justified asymptotically and is reasonably accurate already at
small values of $n$,
we perform these integrals and get
\begin{equation}
\frac{8\alpha_s^2}{9\pi^3}
\left( 2\mu \frac{9 \zeta_2^2}{16} \ln \frac{9 T^2}{m_q^2} +
2 \delta T\frac{3\zeta_2}{4} \frac{7 \zeta_3}{4}
\ln \frac{15 T^2}{m_q^2} \right) T
\end{equation}
The integrals over $ E_p d^3p$ and $p_z d^3 p$ are performed similarly and
introduce no further complications.

\centerline{\bf Scattering Processes}

Scattering diagrams are somewhat more complicated because
the sum over chemical potentials and energies involves both incoming and
outgoing particles.  Consider the $t$ channel gluon exchange diagrams.
The amplitude squared and summed over outgoing states is
$\simeq 40g_s^4s^2/t^2$ for a top quark scattering off a gluon and
$=(5/6)32g_s^4(s^2+u^2)/t^2\simeq (160/3) g_s^4s^2/t^2$
for a top quark scattering off a quark (when it scatters from another
top quark, $\sum \delta = 0$ as the other top is at the same temperature
and velocity).  We have replaced $u\rightarrow -s$
which is legitimate in the leading log approximation.
The sum over perturbations $\sum \delta = 0 \mu +
(E_p - E_{p'}) \delta T + (p_z-p_z') v$ so these diagrams do not damp chemical
potential.  Also, the integration
measure is symmetric under $p \leftrightarrow p'$ and
$k \leftrightarrow k'$, but the integrand is
antisymmetric, so the diagrams do not contribute to the first fluid equation.
Also, the integral $p_z d^3p$ gets no contribution from $\delta T$ because
the integrand is invariant under parity transformation
$\vec p_i\rightarrow -\vec p_i$,
with $\vec p_i=\{ \vec p, \vec k,\vec p^{\,\prime},\vec k^{\prime}\}$
and the integral over $E_p d^3p$ similarly
gets no contribution from $v$.  Hence we need only two integrals,
\begin{equation}
\int_{pkp'k'} \frac{A}{T^4} \frac{s^2}{(t-m_g^2)^2} (2\pi)^4 \delta(p+k-p'-k')
f_pf_k(1-f_{p'})(1 \pm f_{k'})\left\{ \begin{array}{l} E_p
(E_p - E_{p'}) \\ p_z(p_z-p'_z)    \end{array} \right.
\end{equation}
where $A=40g_s^4$ for the scattering off a  gluon, and $A=(160/3) g_s^4$
for the scattering off a quark and of course the
correct population densities for bosons and fermions ought to be chosen.
We use the symmetry of the integration measure and ${\cal P}[f]$ under
$p \leftrightarrow p'$ to rewrite $E_p(E_p - E_{p'}) \rightarrow (E_p -
E_{p'})^2 /2$ and write it in a Lorentz invariant form as $[u \cdot (p-p')]^2
/2$, where $u_{\mu}$ is the unit vector in the time direction of
the plasma frame.  Similarly, $p_z(p_z-p_{z}') \rightarrow -t/6 +
[u \cdot (p-p')]^2 / 6$.

Writing the center of mass frame angle between $\vec{u}$ and $\vec{p}$ as
$\beta$, the integrals over $p'$ and $k'$ give, in the center of mass frame,

\begin{equation}
{A\over 8\pi}\left \{ \begin{array}{c} (|\vec{u}|\, |\vec{p}| \sin\beta)^2
\ln{4p^2\over m_g^2} \\
\frac{1}{3}\left [ (|\vec{u}|\, |\vec{p}| \sin\beta)^2
 +2p^2\right ]\ln{4p^2\over m_g^2} \end{array}\right.
\label{version1}
\end{equation}
A quick calculation gives
\[
|\vec{u}|_{cm} |\vec{p}|_{cm} \cos \beta = \vec{u} \cdot \vec{p}
= u \cdot (p-k) / 2 =  \frac{(E_p - E_k)_{\rm plasma}}{2}
\]
and
\[
 |\vec{u}|_{cm} |\vec{p}|_{cm} = \frac{ |\vec{p} + \vec{k}|_{\rm plasma}}{2}
\]
so, in terms of the plasma frame angle $\theta$ between $\vec{p}$ and
$\vec{k}$, (\ref{version1}) is
\begin{equation}
\frac{A}{16 \pi} \left \{ \begin{array}{c} E_p E_k (1 + \cos \theta)
\ln (2E_p E_k (1-\cos\theta)/m_g^2) \\  \frac{1}{3} E_p E_k (3 - \cos \theta)
 \ln (2E_p E_k (1-\cos\theta)/m_g^2) \end{array} \right.
\end{equation}
Recall that $A=40 g_s^4\; (A=(160/3) g_s^4)$ for scattering off a gluon
(quark). We can now do the remaining integrations as above, using the same
leading log approximation as in the previous section.
the result is
\begin{eqnarray}
\frac{5}{\pi^3} 2\zeta_2^2
\alpha_s^2\ln{25 T^2\over m_g^2} T
\left \{ \begin{array}{c} 1/2\\ 1/2 \end{array}
\right.
\,,\qquad
\frac{20}{3\pi^3} \zeta_2^2
\alpha_s^2\ln{25 T^2\over m_g^2}  T
\left \{ \begin{array}{c} 1/2\\ 1/2 \end{array}
\right.
\end{eqnarray}
for scattering off a gluon and a quark, respectively.  Bose statistics have
made scattering from a gluon more important than from a quark, even though
the matrix element squared is smaller.

\centerline{\bf Results for top quark diagrams}

We now tabulate the contributions of the collision integrals
to the decay constants $\Gamma$ for the top quarks.

The contributions of two annihilation diagrams, both with
matrix elements $\simeq -(64/9)st/(t-m_q^2)^2$, sum to
\begin{eqnarray}
\frac{32\alpha_s^2}{9\pi^3}T \left\{ \begin{array}{ll}
({9\zeta_2^2/16})\ln(9T^2/m_q^2) & \Gamma_{\mu1} \\
({21\zeta_2\zeta_3}/{16})\ln(15T^2/m_q^2) & \Gamma_{\mu2} \\
({21\zeta_2\zeta_3}/{16})\ln(15T^2/m_q^2) & \Gamma_{T1} \\
({135\zeta_2\zeta_4}/{64})\ln(21T^2/m_q^2)
 + ({49 \zeta_3^2}/{32}) \ln(25T^2/m_q^2) & \Gamma_{T2} \\
({45\zeta_2\zeta_4}/{64})\ln(21T^2/m_q^2) & \Gamma_{v} \end{array} \right.
\label{annihilation}
\end{eqnarray}

The scattering diagrams when summed yield
\begin{eqnarray}
\frac{25\alpha_s^2}{3\pi^3} T \left\{ \begin{array}{ll}
\zeta_2^2 \ln(25T^2/m_g^2) & \Gamma_{T2} \\
\zeta_2^2  \ln(25T^2/m_g^2) & \Gamma_{v}
\end{array} \right.
\end{eqnarray}

Absorption and re-emission of a gluon,
with matrix element $\simeq -(64/9) st/(t-m_q^2)^2$, contributes
\begin{eqnarray}
\frac{\alpha_s^2}{36\pi^3} T \left\{ \begin{array}{ll}
135\zeta_2\zeta_4  \ln(21T^2/m_q^2)
- 98 \zeta_3^2 \ln(25T^2/m_q^2)
& \Gamma_{T2} \\
90\zeta_2\zeta_4 \ln(21T^2/m_q^2)
- 98 \zeta_3^2 /3 \ln(25T^2/m_q^2)
 & \Gamma_{v}
\end{array} \right.
\label{absorbtionreemission}
\end{eqnarray}

These collision integrals were evaluated in analogous manner to the gluon
exchange scatterings; after integrating over $p'$ and $k'$ one obtains
$(16g_s^2/9\pi)(\gamma v p \cos\beta)^2\rightarrow (16g_s^2/9\pi)
(1/4) [ (p-k')\cdot u ]^2$ for the contribution to $\delta T$ and
$(16g_s^2/27\pi)[(\gamma v p \cos\beta)^2+p^2]\rightarrow (16g_s^2/27\pi)
(1/4)\{[ (p-k')\cdot u ]^2+2p\cdot k \}$ for the contribution to $v$.

Evaluating (\ref{annihilation}) -- (\ref{absorbtionreemission})
numerically, using $\alpha_s = 0.12$, $m_g^2 = 2g_s^2 T^2$,
and $m_q^2 = g_s^2 T^2/6$, we find
$\Gamma_{\mu1f}=.008993 T$,
$\Gamma_{\mu2f}= .01752 T = T\Gamma_{T1f}$,
$\Gamma_{T2f}= 0.07172 T$,
and $\Gamma_{vf}= 0.03765 T$ \cite{onpaper1}.
By comparing the contributions from various diagrams one finds that
scatterings dominate $\Gamma_v$, but $\Gamma_{T2}$ arises mainly from
annihilations.

We should comment that we have left out one potentially important diagram,
weak flavor changing scattering, which converts left handed top quarks to
bottom quarks.  Because of the linear Coulomb singularity, cut off by Debye
screening, this diagram contributes at order $\alpha_w$, and numerical
evaluation shows that its contribution to the decay rate of left handed
top quarks is comparable to that from the annihilation processes we have
considered.  However, it only affects left handed top quarks, and even
if the rate were infinite it would just share their chemical potential
equally with left handed bottom quarks, reducing the friction on the wall
by 3/4.  (The rate at which thermal top quarks rotate between right and left
handed on the wall is smaller than the annihilation rate.)  We will neglect
this fairly minor effect here.

\centerline{\bf W boson diagrams}

For the $W$ bosons the dominant annihilation processes are $t$ channel
conversion to a gluon by a quark and $W$ -- gluon fusion to a quark-antiquark
pair.  Summing over generations, flavors, colors, and particle-antiparticle,
we find the matrix element is
$\simeq -24g_s^2g_w^2 st/(t-m^2)^2$ for each process.
(Again we only consider the leading log so $s \simeq -u$.)
We will also include order $\alpha_w^2$ processes, which are double $W$
fusion to fermions, with matrix element $\simeq -18st/(t-m^2)^2$,
$W$ scattering from a fermion, with matrix
element $120g_w^4 s^2/(t-m_W^2)^2$, and absorption re-emission, with matrix
element $\simeq -18 g_w^4 st/(t-m^2)^2$.  Scattering from another $W$ boson
does not contribute to the decay rates we consider because the sum of particle
number, $E$, and $\vec{p}$ over all particles is zero.  These processes do
contribute to the damping of higher order perturbations, however, which may
help to thermalize infrared $W$ bosons.

We will neglect annihilation to and scattering from the Higgs doublet because
the matrix elements are more complicated, because the diagrams introduce
infrared problems, and finally and most importantly because there is only
one doublet, so its contribution
is much smaller than the 12 fermion doublets.

The collision integrals are completely analogous to those discussed above, so
we will only present the results.  For semi-strong annihilation, we find
\begin{equation}
\frac{6}{\pi^3} \alpha_s \alpha_w T \left\{ \begin{array}{ll}
({9\zeta_2^2}/{16})\ln(9T^2/m_q^2) & \Gamma_{\mu1} \\
({21\zeta_2\zeta_3}/{16})\ln(15T^2/m_q^2) & \Gamma_{\mu2} \\
({21\zeta_2\zeta_3}/{16})\ln(15T^2/m_q^2) & \Gamma_{T1} \\
({135\zeta_2\zeta_4}/{32})\ln(21T^2/m_q^2)
& \Gamma_{T2} \\
({45\zeta_2\zeta_4}/{32})\ln(21T^2/m_q^2) & \Gamma_{v} \end{array} \right.
\end{equation}
For doubly weak annihilation, we find
\begin{equation}
\frac{9}{2\pi^3} \alpha_w^2 T \left\{ \begin{array}{ll}
({9\zeta_2^2}/{16}) \ln(9T^2/{\langle m^2 \rangle}) & \Gamma_{\mu1} \\
({21\zeta_2\zeta_3}/{16})\ln(15T^2/ {\langle m^2 \rangle})
 & \Gamma_{\mu2} \\
({21\zeta_2\zeta_3}/{16}) \ln(15T^2/ {\langle m^2 \rangle})
 & \Gamma_{T1} \\
({135\zeta_2\zeta_4}/{64}) \ln(21T^2/{\langle m^2 \rangle})
+ ({49 \zeta_3^2}/{32}) \ln(25 T^2/{\langle m^2 \rangle})
& \Gamma_{T2} \\
({45\zeta_2\zeta_4}/{64}) \ln(21T^2/{\langle m^2 \rangle})
 & \Gamma_{v} \end{array} \right.
\end{equation}
where $\ln({\langle m^2 \rangle})= 3\ln(m_q^2)/4 + \ln(m_l^2)/4$.

For collisions from quarks and leptons, we find
\begin{equation}
\frac{15}{2\pi^3} \alpha_w^2 T \left \{ \begin{array}{ll}
2\zeta_2^2 \ln ({25 T^2}/{ m_W^2}) & \Gamma_{T2} \\
2\zeta_2^2 \ln ({25 T^2}/{m_W^2})  & \Gamma_{v}
\end{array}\right.
\end{equation}
and just as in the case of quarks, scattering diagrams contribute equally to
the $\Gamma_{T2}$ and $\Gamma_v$.
For absorption and re-emission from fermions, we find
\begin{equation}
\frac{9}{4\pi^3} \alpha_w^2 T \left \{ \begin{array}{ll}
({135\zeta_2\zeta_3}/{32})\ln ({21 T^2}/{ \langle m^2\rangle } )
-({49\zeta_3^2}/{16})\ln ({25 T^2}/{\langle m^2\rangle })
& \Gamma_{T2} \\
({45\zeta_2\zeta_3}/{16})\ln ({21 T^2}/{\langle m^2\rangle })
- ({49\zeta_3^2}/{48})\ln ({25 T^2}/{\langle m^2\rangle } )
& \Gamma_{v}
\end{array}\right.
\end{equation}

The thermal mass of a left-handed lepton is $m_l^2 = 3g_w^2 T^2/32$ plus a
small hypercharge correction which we neglect.
The results, using $\alpha_w = 1/30$, are
$\Gamma_{\mu1} = 0.00521 T$,
$\Gamma_{\mu2} = 0.01012 T = \Gamma_{T1}$,
$\Gamma_{T2} = 0.03686 T$,
and $\Gamma_{v} =  0.01614 T$.  Annihilations dominate even $\Gamma_v$.

\section{Appendix B:  Critique of the Fluid Approximation}

Is the fluid approximation any good?

The discussion can be broken into two parts: does the fluid approximation
model properly the energy dependence of $\delta f$; and does it oversimplify
the anisotropy of $\delta f$?

We begin with the energy dependence, using the tools developed in Appendix A.
Note first that top quarks decay fairly quickly (we have seen that the tails
around a source have a length of about $5/T$),
so a particle excess decays before it is transported off the
wall.  Hence the friction from one excess particle is roughly its force on
the Higgs field at the point where it was created times its lifetime,
$\propto \tau/E$.  We must determine the behavior of $\tau$.

The annihilation rate of one particle can be computed by putting a bump in
the population function $f$ at a specific energy and examining the collision
integral.  The contribution to
$\delta$ due to the bump is $-1/f_0'$ times the bump,
and $f_0' = - f_0 (1 - f_0)$ for a fermion,
so the population factor in the collision
integral is $f_2 (1 \pm f_3) (1 \pm f_4) / (1 - f_1)$ times the bump.  The
full energy dependence of the collision integral is this expression, the
energy conserving delta function, and the $1/E_p$ prefactor.  The argument is
that $1 - f_1$ is quite weakly energy dependent (going from 1/2 at $E=0$ to
1 at large $E$ where most of the particles are) and for thermal outgoing
particles the integral over outgoing states depends on $E$ only through
$\ln(2p \cdot k/ m_g^2)$, a weak dependence.  The dominant dependence is then
the $1/E$ prefactor, which cancels the $1/E$ strength of the particle's
influence on the Higgs field.  The argument is strongest for thermal particles,
but as they dominate the total friction from top quarks, this should
suffice.

Let us check the performance of the fluid approximation for top quarks
by neglecting spatial derivatives (ie assuming the
particles really do not leave the wall) and comparing the friction from
the fluid equations to the friction we would find if we had only included
a $\mu$ term.  Including only a $\mu$ term, we find
\[
\mu \Gamma_{\mu1} = v_w c_1 m m' \qquad
{\rm friction} = c_1 \mu m m'
\]
The fluid approximation gives
\[
\left[ \begin{array}{l} \mu \\ \delta T \end{array} \right] = \frac{v_w m m'}
{\Gamma_{\mu1} \Gamma_{T2} - \Gamma_{\mu2} \Gamma_{T1}}
\left[ \begin{array}{lll} c_1 \Gamma_{T2} & - & c_2 \Gamma_{T1} \\
c_2 \Gamma_{\mu 1} & - & c_1 \Gamma_{\mu2} \end{array} \right]
\qquad
{\rm friction} = (c_1 \mu + c_2 \delta T) m m'
\]
Plugging in the values for the $\Gamma$'s, the friction is $v_w m^2 m'^2$
times $.137$ in the first case and $.143$ in the second case; using the full
fluid approximation has only changed our estimate of the friction
by 4\%.  (If the relaxation time approximation were valid the value
would have been different by 27\%, so the argument must have some
validity.)
Since we have difficulty trusting our evaluation of the collision
integrals to better than 50\% (mainly because of the leading log
approximation), and since the fluid equations we use should
account reasonably well for transport, there seems little reason to improve
the fluid approximation for top quarks, except by improving the precision
of our evaluation of the collision integrals.

The argument for the fluid approximation
also works for thermal $W$ bosons, where $1 + f \simeq 1$; but
for soft $W$ bosons, the population term in the collision integral is
$f_2 (1 \pm f_3)(1 \pm f_4)/(1 + f_1) \simeq f_2 (1 \pm f_3)(1 \pm f_4) E/T$,
which cancels the $1/E$ in front of the fluid equations; apparently,
the decay rate does not rise as we lower the energy.
Evaluating the annihilation and
scattering rates of the infrared $W$ bosons is further confused by the
appearance of loop diagrams (hard thermal loops) which enter at the same
parametric order as tree level effects.  Also we should include $W$ boson
scattering from other $W$ bosons, which does not contribute to the decay
rates we have computed but does contribute to the rate at which infrared
particles are thermalized.  We will not attempt to treat this problem here,
but will only remark that the fluid approximation appears to be a very
naive treatment of the $W$ bosons.

Next, we will discuss whether it is acceptable to treat the anisotropy of the
distribution keeping only the lowest term, $v \cdot p = p v_i Y_{1i}(\hat{p})$,
 rather
than including higher angular moments $p^l Y_{lm}(\hat{p})$, where $Y_{lm}$
is a spherical harmonic.  The source of $\delta f$ is isotropic,
so $\vec{v}$ only arises out of spatial gradients of $\delta f$.  The decay
of $\vec{v}$ is characterized by the diffusion length $D \simeq 2.7/T$
for quarks (and $5.5/T$ for $W$ bosons), which is generally
smaller than the thickness of the wall.  Relative to the other perturbations
its amplitude is down by $D/L$.  We expect perturbations at high angular
moments
to be down by $(D/L)^l$; as long as $D$ is sufficiently less than $L$, we
expect them to be less important than $\vec{v}$.  To confirm this we
can extend our treatment beyond the fluid approximation to account for rank
two tensor deviations.  Setting $T=1$ for simplicity (we can restore it by
dimension counting) and writing
\begin{eqnarray}
f^{-1}  & = &  \exp( E - \delta) \pm 1   \nonumber \\
 \delta  & = & \mu + E \delta T + p_i v_i + E^2 \epsilon + E p_i \epsilon_i
+ (p_i p_j - \delta_{ij} p^2/3) \epsilon_{ij}
\nonumber
\end{eqnarray}
we find extended fluid equations;
\begin{eqnarray}
c_2 \dot{\mu} +c_3 \delta \dot{T} +c_4 \dot{\epsilon} +
\frac{c_3}{3} \partial_i v_i +\frac{c_4}{3} \partial_i \epsilon_i
-c_1 m\dot{m}  =  - \Gamma_{\mu1} \mu
- \Gamma_{T1}\delta T - \Gamma_{\epsilon1}\epsilon
\label{ex1} \\
c_3 \dot{\mu} +c_4 \delta \dot{T} +c_5 \dot{\epsilon} +
\frac{c_4}{3} \partial_i v_i +\frac{c_5}{3} \partial_i \epsilon_i
-c_2 m\dot{m}  =  - \Gamma_{\mu2} \mu
- \Gamma_{T2}\delta T - \Gamma_{\epsilon2}\epsilon
\label{ex2} \\
c_4 \dot{\mu} +c_5 \delta \dot{T} +c_6 \dot{\epsilon} +
\frac{c_5}{3} \partial_i v_i +\frac{c_6}{3} \partial_i \epsilon_i
-c_3 m\dot{m}  =  - \Gamma_{\mu3} \mu
- \Gamma_{T3}\delta T - \Gamma_{\epsilon3}\epsilon
\label{ex3} \\
\frac{c_3}{3}\partial_i\mu + \frac{c_4}{3}\partial_i\delta T +
\frac{c_5}{3}\partial_i \epsilon + \frac{c_4}{3}\dot{v}_i +
\frac{c_5}{3} \dot{\epsilon}_{i} + \frac{c_5}{15}( \partial_j \epsilon_{ij}
+ \partial_j \epsilon_{ji} - \frac{2}{3} \partial_i \epsilon_{jj})
\qquad \nonumber \\
 =  - \Gamma_{v1} v - \Gamma_{\epsilon i1}\epsilon_i
\label{ex4} \\
\frac{c_4}{3}\partial_i\mu + \frac{c_5}{3}\partial_i\delta T +
\frac{c_6}{3}\partial_i \epsilon + \frac{c_5}{3}\dot{v}_i +
\frac{c_6}{3} \dot{\epsilon}_{i} + \frac{c_6}{15}( \partial_j \epsilon_{ij}
+ \partial_j \epsilon_{ji} - \frac{2}{3} \partial_i \epsilon_{jj})
\qquad \nonumber \\
 =  - \Gamma_{v2} v - \Gamma_{\epsilon i2}\epsilon_i
\label{ex5} \\
\frac{c_5}{15} ( \partial_i v_j + \partial_j v_i - \frac{2 \delta_{ij}}{3}
\partial_k v_k ) +
\frac{c_6}{15} ( \partial_i \epsilon_j + \partial_j \epsilon_i
- \frac{2 \delta_{ij}}{3} \partial_k \epsilon_k )
 + \nonumber \\
\frac{c_6}{15}( \dot{\epsilon}_{ij} + \dot{\epsilon}_{ji}
- \frac{2 \delta_{ij}}{3} \dot{\epsilon}_{kk})
  =  - \Gamma_{\epsilon ij} \epsilon_{ij}
\label{ex6}
\end{eqnarray}
The matrix of $\Gamma$'s is block diagonal in angular moments
and symmetric.  It contains 10 distinct
nonzero terms; its evaluation is the main impediment to including higher
tensor moments.

Only the isotropic moments $\mu$, $\delta T$, and $\epsilon$ are directly
sourced, since only their time derivatives appear in the equations with
the source terms $c m \dot{m}$.
The vector moments $v_i$ and $\epsilon_i$ are sourced by gradients
in the isotropic moments.  We expect their amplitudes to be $\sim \mu'/\Gamma
\sim \mu D/L$.  Similarly, the traceless tensor moment $\epsilon_{ij}$ is
sourced by gradients in the vector moments, $\partial_i v_j$.  Although we
have not calculated $\Gamma_{\epsilon ij}$,
we anticipate that it will be at least
as large as the damping rates for the vector moments, because a particle's
contribution to $\epsilon_{ij}$ is erased by deflecting the particle
by $45^{\circ}$, while it must be deflected by $90^{\circ}$ to erase its
contribution to $v_i$.  Hence, $\epsilon_{ij}$ should be $\sim v_i D /L
\sim \mu D^2 / L^2$.  We conclude that as long as $L >> D$, the neglect of
high order angular moments should be justified.  (In the opposite limit,
$D >> L$, we know that, for quarks,
the high order moments are very important and lead to a
log enhancement of the friction; so $L >>D$ is not only a sufficient
condition to neglect high angular moments, but a necessary one.)  In our
case, $D \simeq 2.7/T$ for top quarks, and we are in good shape.
(Even for $W$ bosons, where $D \simeq 5.5/T$, we are in fairly good shape.)

We wish to note that this conclusion relies on the fact that we are interested
in the motion of one fluid against another, and not in the dissipation of
the motion of a fluid as a whole, or of a fluid of one species.  Let us
explore this case briefly, to understand the physics of the extended fluid
equations.  When tracking a one component fluid, or the average over
species of a several component fluid,
energy and momentum conservation ensure that
$\Gamma_{v1}$, $\Gamma_{T2}$, $\Gamma_{T1} = \Gamma_{\mu2}$,
 $\Gamma_{v2}=\Gamma_{\epsilon i1}$, and
$\Gamma_{T3} = \Gamma_{\epsilon 2}$
 vanish identically, so $\delta T$ and $v_i$
are not directly dissipated.  If in addition there is a nonzero conserved
charge density for which $\mu$ is the chemical potential, then
$\Gamma_{\mu1}$ and
 $\Gamma_{\epsilon1} = \Gamma_{\mu3}$ also vanish, and $\mu$ is not directly
dissipated.  It is still true that $\epsilon_{ij} \sim v_i D/L$, but now its
presence is important, as it is the main source of dissipation in the system.
Indeed, if the perturbations are slowly varying, then we can neglect
$\dot{\epsilon}_{ij}$ compared to $\partial_i v_j$, et cetera, and solve for
the high order perturbations in terms of the lower order ones.  We find
\begin{eqnarray}
\epsilon_{ij} = \frac{- c_5}{15 \Gamma_{\epsilon ij}} ( \partial_i v_j +
\partial_j v_i -\frac{2 \delta_{ij}}{3} \partial_k v_k) \nonumber \\
\epsilon_i = \frac{-1}{3 \Gamma_{\epsilon i2}} (c_5 \dot{v}_i + c_5
\partial_i \delta T + c_4 \partial_i \mu)  = \frac{-1}{3 \Gamma_{\epsilon i2}}
(c_4 - \frac{c_3 c_5 }{c_4} ) \partial_i \mu
\nonumber \\
\epsilon = \frac{-1}{\Gamma_{\epsilon 3}}(\frac{c_5}{3} \partial_i v_i
+ c_5 \delta \dot{T} + c_4 \dot{\mu})  \nonumber
\end{eqnarray}
all plus $O(\delta ''/\Gamma^2)$.  Here we have used Eq. (\ref{ex4}) in the
equation for $\epsilon_i$.  We can also use
Eqs. (\ref{ex1}) and (\ref{ex2}) to
simplify the relation for $\epsilon$, but here we should note that
the coefficients on $\partial_i v_i$ and $\delta \dot{T}$ only equal in the
massless limit.  If the mass is zero, then $\epsilon=0$;
otherwise it is $- O(m^2) \partial_i v_i /\Gamma_{\epsilon 3}$.

Substituting these three quantities into the fluid equations for the
undamped species, we find
\begin{eqnarray}
c_2 \dot{\mu} + c_3 \delta \dot{T} + \frac{c_3}{3} \partial_i v_i  & = &
c_1 m \dot{m} + \frac{c_4^2 - c_3 c_5}{9 \Gamma_{\epsilon i2}} \partial^2 \mu
+ \frac{O(m^2)}{\Gamma_{\epsilon 3}}  \partial_i \dot{v}_i \nonumber \\
c_3 \dot{\mu} + c_4 \delta \dot{T} + \frac{c_4}{3} \partial_i v_i  & = &
c_2 m \dot{m} + \frac{c_4^2 c_5 - c_3 c_5^2}{9 c_4 \Gamma_{\epsilon i2}}
\partial^2 \mu
+ \frac{O(m^2)}{\Gamma_{\epsilon 3}}  \partial_i \dot{v}_i \nonumber \\
\frac{c_3}{3} \partial_i \mu + \frac{c_4}{3} \partial_i \delta T +
\frac{c_4}{3} \dot{v}_i & = &  \frac{c_4^2 c_5 - c_3 c_5^2}
{9 c_4 \Gamma_{\epsilon i2}} \partial_i \dot{\mu}
+ \frac{O(m^2)}{\Gamma_{\epsilon 3}}  \partial_i \partial_j v_j
 \nonumber \\ & &
+ \frac{2 c_5^2}{225 \Gamma_{\epsilon ij}} \partial_j (\partial_i v_j +
\partial_j v_i - \frac{2 \delta_{ij}}{3} \partial_k v_k)
\end{eqnarray}
which are the
linearized relativistic Navier-Stokes equations.  The perturbations
$\epsilon$, $\epsilon_i$, and $\epsilon_{ij}$ have caused bulk viscosity,
thermal resistivity, and shear viscosity, respectively.  The bulk viscosity
vanishes in the ultrarelativistic ($m << T$) limit; the thermal resistivity
vanishes in the absence of a nonzero conserved charge density.  (In this
case heat flow is only resisted by higher derivative terms.)

\section{Appendix C:  Runaway Wall}

According to the tanh {\it Ansatz}, the wall is capable of becoming
ultrarelativistic without contracting to a small plasma frame thickness.  We
will explore how this arises and why it is unphysical.

First consider the equation of motion in the tanh {\it Ansatz}, Eq.
(\ref{constraints}) and particularly Eq. (\ref{eq:constraint two}).
We see that the derivative term $\Box \phi$ stretches the wall, but that
it becomes ineffective at large velocity.  One would expect, then, that
the wall becomes extremely compressed.  But the frictive term
Eq. (\ref{horrorhorror2}) stretches the wall.  For a thin wall, where the
particles have little time to decay, and at large velocity, where they
simply sweep backwards up the wall, we find $\mu \propto \phi^2$ and the
ratio of stretching to friction is $11/12$.  When $\Xi < 5\Delta V_T/6$,
then if there is enough friction to stop the wall from accelerating then
there is enough to prevent it from contracting to the regime where we
can neglect particle decay, so the wall cannot become thinner than about
$L \sim 6/T$.  When $A_F \neq 0$, the system tends to supercool quite
heavily and $\Xi$ tends to be small compared to $\Delta V_T$; the wall
is then susceptible to runaway, according to the tanh {\it Ansatz}.

Note that most of the frictive force on the wall comes on the upper part
near the symmetric phase.  The friction tends to make this part of the
wall very thick, as we see explicitly when we solve for the wall shape.
What the tanh {\it Ansatz} does is force the front part of the wall to be
as thick as the back, which is unphysical.

Let us attempt an analysis without a wall shape {\it Ansatz}.
Consider the wall propagating at a steady,
very relativistic speed, say $\gamma v_w
> 10$.  To determine how abrupt the wall is, we will integrate the
equation of motion times $\phi'$, starting in the symmetric phase
and going up to the point $\phi = \phi_1$; if the wall is in a steady state,
\begin{equation}
\int_{-\infty}^{z : \phi = \phi_1}( V_T'(\phi(z)) + {\rm friction})\phi'
 \; dz = (1-v_w^2) \int \phi' \phi'' dz = (1-v_w^2) \frac{(\phi')^2}{2}
\end{equation}

Because $1-v_w^2$ is very small, the wall is rising very rapidly where
$\phi = \phi_1$
unless $V_T(\phi_1) + \int {\rm friction}\phi'\, dz$ is almost zero.
Let us examine whether this condition is ever satisfied.

First note that the friction is never negative anywhere for a monotonic wall.
Now $V_T$ is positive at small $\phi$, as otherwise the phase transition would
have already proceeded by spinodal decomposition.  So there will be a
section at the front of the wall where $\phi'$ is large, ie a section which
is very abrupt.  To determine where this section ends, we need to find
$\int {\rm friction}\phi' \, dz$ on this section of wall.

This is easy in the fluid approximation.  As the wall is thin, we can neglect
decays, and the fluid equations can be integrated.  The friction is, in
the notation of section \ref{ansatzz}, $f_i (A^{-1})_{ij}F_j \phi_1^4/8$.
Adding this to the effective potential, we get a curve whose second minimum
lies above 0 for the effective potential parameters in
Table \ref{velocitytable} with
$A_F = 0.1$ or 0.2, but whose second minimum is below 0 for $A_F = 0.3$.
In the former case, the whole wall is abrupt and feels a net backwards
force; the wall will then slow down until it is not abrupt and no runaway
occurs.  In the latter case, a section of the wall feels a net forward
pressure, and accelerates without limit.  According to the fluid approximation,
the wall will run away in these cases.

When the wall becomes this thin,
the fluid approximation is definitely
inaccurate--as we have argued, it only makes sense when the wall is
thicker than the diffusion length.  The free particle approximation
should be appropriate, however, and for an ultrarelativistic wall we can
make an accurate expansion in large $\gamma$.

The total pressure on this section of wall from one
species is, in the free particle, 1 loop approximation (\cite{DineLinde,Mac})
\begin{eqnarray}
\int_m^\infty \frac{dp_z}{2\pi} (p_z - \sqrt{p_z^2 - m^2})\int
\frac{d^2 p_\bot}{(2\pi)^2} \frac{p_z}{\sqrt{p_z^2 + p_\bot^2}} \frac{1}
{\exp(\beta \gamma(p - vp_z)) \pm 1} +
\nonumber \\
\int_0^m \frac{dp_z}{2\pi} 2p_z \int
\frac{d^2 p_\bot}{(2\pi)^2} \frac{p_z}{\sqrt{p_z^2 + p_\bot^2}} \frac{1}
{\exp(\beta \gamma(p - vp_z)) \pm 1} +
\quad \\
\int_0^\infty \frac{dp_z}{2\pi} (-p_z + \sqrt{p_z^2 + m^2})\int
\frac{d^2 p_\bot}{(2\pi)^2} \frac{p_z}{\sqrt{p_z^2 + p_\bot^2 + m^2}} \frac{1}
{\exp(\beta \gamma(\sqrt{p^2 + m^2} + vp_z)) \pm 1} \nonumber
\end{eqnarray}
where $m$ is the particle mass when $\phi = \phi_1$.

Taking $\gamma >> T/m$ so that $\exp(-\gamma m/T) << m^2/T^2$,
we find the friction is
\begin{equation}
\frac{m^2 T^2}{48} + \frac{m^3 T \ln 2}{12\pi^2 \gamma}
- \frac{m^2 T^2}{192\gamma^2} + O(m^4 \ln (\gamma T/m) /\gamma^2)
\end{equation}
from a fermion degree of freedom and
\begin{equation}
\frac{m^2T^2}{24} + \frac{m^3T}{4\pi^2\gamma} \left( -\frac{5}{18} - .1347
 + \frac{\ln(2\gamma T/m)}{3} \right) - \frac{m^2 T^2}{96\gamma^2}
+O(m^4 \ln (\gamma T/m) / \gamma^2 )
\end{equation}
from a boson\footnote{.1347097 is the numerical value of $\int_1^\infty
(x^2 - x\sqrt{x^2 -1} -1/2) \ln x \, dx$, which we reduced to $-\sum_{n=2}
\Gamma(n-1/2)/( \Gamma(-1/2) \Gamma(n+1) (2n-3)^2)$ but
were unable to reduce further.}.
Subtracting from these the corresponding 1 loop
contributions to the effective potential,
\begin{equation}
\frac{m^2T^2}{48} - \frac{m^4}{32\pi^2}\left( \ln( \pi T/m ) -  \gamma_E
+ .75 \right) \, , \qquad \frac{m^2 T^2}{24} - \frac{m^3 T}{12\pi}
\end{equation}
gives the friction from particles on the abrupt part of the wall, at 1 loop.
This friction is much larger than the prediction of the fluid approximation.
Adding this expression to $V_T$ produces a function which is positive for
all $\phi_1 > 0$, for all values of the effective potential parameters we
have considered; in fact it comes quite far from having a second zero.
This means that there is no value of $\phi$ where the abrupt part of the wall
can end; it always has a net backwards force on it, even when
$\phi_1 = \phi_0$.  It is simply impossible to have an ultrarelativistic,
steady state solution to the wall shape.  As the wall becomes very fast, the
front of the wall always compresses and sustains enough friction to prevent
further acceleration.  We
cannot determine the velocity of the wall, but we can say that
it must be slow enough that the free particle approximation is inaccurate,
which ensures that $\gamma v_w$ cannot be much greater than 1.
Note that it was not necessary for the condensate responsible for $A_F$ to
exert any friction for this to be true.

\pagebreak

\begin{figure}
\caption{All annihilation, scattering, and absorption re-emission diagrams
considered in the text}
\label{diagrams}
\end{figure}

\begin{figure}
\caption{Plot of the solutions of Eq. (\protect\ref{constraints}) for the case
$\lambda_T = 0.03$, $A_F = 0$.  The solid line is the velocity
constraint and the dotted line is the thickness constraint.}
\label{vLmatch}
\end{figure}

\begin{figure}
\caption{Wall shape; at left, for $\lambda_T = 0.04$, $A_F=0.1$, and at right,
for $\lambda_T = 0.03$, $A_F = 0$.  The solid line is
the numerical value, the dotted line is the tanh {\it Ansatz} shape.
The scale is the same for both; the full width of the pictures are $96/T$
and $64/T$.}
\label{Blah}
\end{figure}


\begin{thebibliography}{99}
\bibitem{onpaper1} these values differ slightly from those in \cite{paper1}
 because there we used a slightly different approximation in Eq. (
\ref{leadinglog}), evaluated the scattering integrals numerically rather
than analytically, and made a slight error in the absorption-reemission
diagram.  The difference is insignificant.
\end{thebibliography}
\end{document}